\documentclass[11pt,a4paper]{article}      
\usepackage{geometry}
\usepackage[affil-it]{authblk}
\usepackage{graphicx}
\usepackage{algorithm,algorithmic}
\usepackage{amsmath}
\usepackage{booktabs}
\usepackage{natbib}
\usepackage[export]{adjustbox}

\let\tt\texttt

\begin{document}

\title{\bf Adaptive large neighborhood search for a personnel task scheduling problem with task selection and parallel task assignments
}

\author[]{Martin Gutjahr}
\author[]{Sophie N. Parragh\thanks{sophie.parragh@jku.at}}
\author[]{Fabien Tricoire}
	
			\affil[]{Institute of Production and Logistics Management/JKU Business School\\
			Johannes Kepler University Linz, Linz, Austria}
			
\date{}

\maketitle

\begin{abstract}

Motivated by a real-world 
application,
we model and solve 
a complex staff scheduling problem. Tasks are to be assigned to workers for supervision. Multiple tasks can be covered in parallel by a single worker, with worker shifts being flexible within availabilities. Each worker has a different skill set, enabling them to cover different tasks. Tasks require assignment according to priority and skill requirements. The objective is to maximize the number of assigned tasks weighted by their priorities, while minimizing assignment penalties. We develop an adaptive large neighborhood search (ALNS)  algorithm, relying on tailored destroy and repair operators. It is tested on benchmark instances derived from real-world data and compared to optimal results obtained by means of a commercial MIP-solver. Furthermore, we analyze the impact of considering three additional alternative objective functions. When applied to large-scale company data, the developed ALNS outperforms the previously applied solution approach.
\end{abstract}

\section{Introduction}
\label{intro}
Motivated by a real-world problem faced by the company Sportradar, which provides betting odds on sports matches, we study a complex personnel scheduling problem. Workers are assigned tasks for supervision. 
Each worker has a specific set of skills that allows them to cover different tasks with varying levels of difficulty. Workers may work on multiple tasks in parallel, with their personal skill determining the limit for parallel tasks. Furthermore, each worker provides their availabilities, within which their shifts can be scheduled. Each worker has requirements for break as well as rest times, depending on the legal requirements at the worker's location. Breaks may be split into parts, however each break incurs a preparation time that is required before resuming work following the break. 
The number of tasks exceeds the 
capacity for coverage. Each task has a given 
priority score  
and asks for a specific skill level in the corresponding task group. The objective is to maximize the total priority score of all assigned tasks, while minimizing the penalty incurred for assigning operators with a skill level lower than the one desired for the assigned tasks. 
We propose a mixed integer linear programming (MIP) formulation for the considered problem and we develop an adaptive large neighborhood search (ALNS) algorithm to solve large-scale instances in a short amount of time. We compare the solutions obtained by the heuristic approach with optimal solutions obtained from solving the developed MIP model 
with CPLEX. Furthermore, the proposed ALNS 
is applied to large-scale real-world data provided by Sportradar. We note that the ideas proposed in this paper  
are not limited to the application to the considered problem, but may 
be used in solution approaches to other  personnel scheduling, rostering and timetabling problems. Their main focus is the requirement of task selection and assignment based on priority scores. Similar problems can be found in aviation and railway traffic like flight cancellations due to general staff shortages or strikes, as well as medical triage. 

The remainder of the paper is structured  
as follows. In Section \ref{rel}, we provide an overview of related work and we present, in Section \ref{def}, 
the  developed mathematical model.  
Section \ref{method} is devoted to the description of the proposed adaptive large neighborhood search algorithm. 
In Sections \ref{results} and \ref{study}, computational results are presented,   
including information on the
proposed test instances, comparison to optimal and to company solutions  
as well as managerial insights derived from changes in the weights of the objective function. Section \ref{conc} concludes the paper and provides possible directions for future research.

\section{Related Work}
\label{rel}
To the best of our knowledge, the problem proposed in this paper has not been considered in the literature before. However, it shares characteristics with a number of well-studied problems. Therefore,
we review the literature in each of these different areas and  point to the main differences and similarities between the reviewed work and the problem considered in this paper.

First and foremost, the problem proposed in this paper belongs to the field of personnel scheduling and rostering problems. With the main goal of the problem being the construction of a schedule for available operators by assigning tasks that already have starting and ending times, the problem is related to 
staff scheduling problems arising in the airline industry, in (public) transportation, or hospitals. 
A main difference between the problem addressed in this paper and those staff scheduling problems 

is the possibility to carry out multiple tasks in parallel. 
\cite{chu2007generating}, e.g., propose a goal programming approach for scheduling and rostering airport personnel and  \cite{maenhout2010hybrid} develop a hybrid scatter search approach for the 
airline crew rostering problem. 

\cite{yunes2005hybrid} consider the crew scheduling and rostering problem in an urban bus transit problem and solve it by means of hybrid column generation approaches. A nurse rostering problem, considering shift assignments and rest times, is, e.g., solved by \citet{solos2013generic} using variable neighborhood search. 
Differing from the problem addressed in this paper, a majority of the research on personnel rostering specifically requires to cover all tasks, with objectives often targeting minimization of (assignment) costs. However, also alternative objectives are addressed, such as, e.g., equity in terms of workload targets \citep{prot2015two}. 

From a methodological point of view, \cite{barrena2013fast} as well as \cite{dong2020integrated} successfully implement ALNS for train scheduling and timetabling problems, while \cite{mansini2019optimizing} use ALNS to schedule physicians in a hospital ward and \cite{sorensen2012international} apply ALNS to high school timetabling.

A subfield of personnel scheduling that is 
relevant for the problem addressed in this paper, is the field of multi-skilling problems, 
where each worker has a set of predefined skills that enable them to carry out a certain subset of the available 
tasks. \cite{cuevas2016mixed}, e.g., propose a mixed integer program for the short-term multi-skilled workforce tour scheduling problem, 
where workers' shifts and days off are planned while assigning activities to shifts.
\cite{de2018three} propose a three stage MIP approach for optimizing skill mix and training schedules for aircraft maintenance personnel. Furthermore, \cite{KPDH10} use ALNS to solve the problem of service technician routing and scheduling to minimize coverage cost, while also taking skill levels of technicians into account.

The problem considered in this paper 
also features the aspect of task selection: 
tasks need to be chosen from a large pool of available tasks with the goal of maximizing the total benefit of covered tasks. While \cite{vijaykumar1998task} already considered the problem of task selection for a multiscalar processor in the last millennium, the field only recently received more attention. \cite{song2010novel}, e.g., consider task selection and allocation for collaborative cloud service platforms and solve the problem using an adaptive filter in combination with a heuristic algorithm. Related problems are often considered in the context of crowd-sourcing. \cite{abououf2019multi} consider multi-worker multi-task selection in mobile crowd sourcing using a genetic as well as a tabu search algorithm, while \cite{deng2016task} consider task selection for spatial crowd-sourcing with the help of dynamic programming and branch-and-bound. \cite{li2018task} propose a game theoretical approach to solve the problem of task selection and scheduling for food delivery and \cite{shaghaghi2017task} consider task selection and scheduling for multifunction radar systems and solve the problem to optimality using branch-and-bound.

Due to the requirement of 
task selection and assignment to build shifts that contain the highest-value tasks, the problem can also be seen 
as a generalization of the bin packing problem.  
Reformulations of scheduling problems as bin packing problems have been proposed going back as far as the 1970s \citep{garey1976resource,coffman1978application}. 
\cite{leinberger1999multi} consider multi-capacity bin packing with application to job scheduling problems and \cite{vijayakumar2013dual} formulate a dual bin packing approach in order to schedule surgical cases in a hospital. More recently, 
\cite{li2018single} have considered a single bin packing machine scheduling problem with two-dimensional bin packing constraints and \cite{witteman2021bin} address  
aircraft maintenance task allocation from a bin packing perspective. Using a hybrid genetic based approach, \cite{su2021hybrid} consider the problem of unrelated parallel workgroup scheduling from a bin packing perspective. Related to the approach considered in this paper, ALNS has been successful at solving bin packing problems. \cite{he2021adaptive} use ALNS to solve the circle bin packing problem and \cite{zeng2021adaptive} propose an ALNS approach for single machine batch processing with two-dimensional bin packing constraints while \cite{zeng2021adaptive2} apply ALNS to a two-dimensional packing problem with conflict penalties.

The problem considered in this paper is also related to resource-constrained project scheduling problems (RCPSPs) in that the assignment of tasks to workers must take capacity restrictions on each worker into account. In contrast to general RCPSPs however, tasks are independent of one another and have fixed starting and ending times. The goal is therefore to use the available resources (worker capacity) as efficiently as possible by choosing the most valuable tasks. Recent research on RCPSPs has, e.g., addressed 
resource leveling 
to keep the use of a scarce resource as balanced as possible \citep{li2018effective} and flexible resource profiles in which resource usage may vary at different times \citep{naber2014mip}. 
Alternative activity chains are, e.g., considered by \cite{tao2017scheduling} and \cite{hauder2020resource}. 
\cite{lova2000multicriteria} address 
RCPSPs with objectives differing from the usual minimization of the makespan as well as multiple projects running in parallel. 
Heuristic approaches related to the approach developed in our paper have been used for RCPSPs to great success, for example by \cite{palpant2004lssper} as well as by \cite{muller2009adaptive,muller2011adaptive}.

Based on the personnel scheduling  
literature review of 
\cite{van2013personnel}, the problem we address falls into the class 
of personnel task scheduling problems, which is defined as follows \citep{krishnamoorthy2001personnel}: a set of tasks with fixed start and end times have to be assigned to a workforce with different qualifications allowing them to carry out only a subset of the available tasks within predefined shifts. In the shift minimization task scheduling problem (SMTSP), the objective is to minimize the total number of employees. The SMTSP relates to list coloring on interval graphs \citep{smet2014shift}, which is NP-complete \citep{bonomo2009exploring}.
We introduce several new aspects 
not considered in the original definition, most important of which is the required task selection.  We therefore refer to the proposed problem as personnel task scheduling problem with task selection (PTSP-TS).
The PTSP-TS generalizes the interval scheduling problem with given machines, where a set of tasks with given start and end times is to be scheduled on a given set of machines and the objective is to maximize the (weighted) number of assigned tasks \citep{kolen2007interval}. The variant where tasks may only be carried out by a subset of the machines (which relates to the skill requirements considered in this paper), is known to be NP-hard \citep{kolen2007interval,arkin1987scheduling}.

\section{Problem Definition}
\label{def}

\begin{figure}
\caption{Graphical display of the assignment of tasks to operators for a toy example} \label{figure:1}
\includegraphics[width=\textwidth, frame]{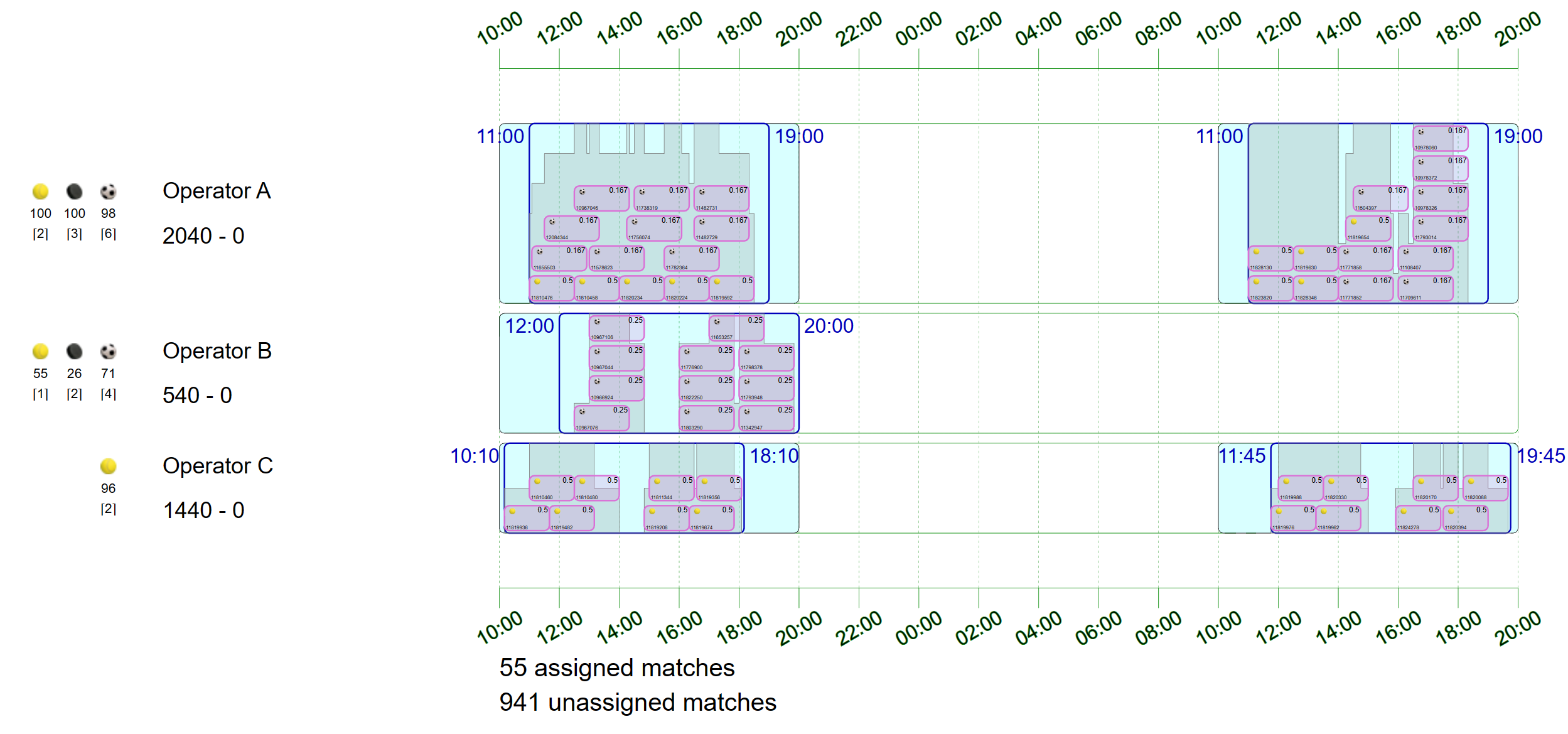}
\centering
\end{figure}

\allowdisplaybreaks
In the PTSP-TS, tasks are to be assigned to operators to maximize the benefit of covered tasks, while minimizing the penalty incurred from assigning lower-skilled operators. In Figure \ref{figure:1}, we show the solution obtained for a toy example using three operators and a total of 996 assignable tasks. Each operator has given skills to cover different tasks. In our application, these skills concern the supervision of different types of sports games.  
In the considered planning horizon of 34 hours, shifts are to be constructed within five availabilities. In the figure, each of these availabilities is represented by a rectangle coloured in teal. The planned shift within the availability is depicted by a blue frame. 
In addition, we give its starting and ending times. Within each shift, each assigned task is represented by a purple rectangle. Its length represents the actual length of the task. The provided floating point value represents the amount of an operator's attention a task requires, later referred to as required bandwidth. The gray bars within a shift give the operator's bandwidth usage. Their heights vary depending on the total bandwidth currently required. Gray bars reaching the top of the shift frame indicate that the bandwidth is fully used. This is, e.g., the case for Operator A with only two tasks in the beginning of the second shift, while six parallel tasks can be feasibly assigned towards the end of the shift.      Below each operator, we display the total assignment score of all the tasks assigned to an operator and the total penalty from assignments below the operator's provided skill level. In this toy example, no tasks below the operators' skill levels are assigned. Therefore, this value is zero. 

In order to more formally state the problem, 
we define a set of tasks $V$ and a set of operators $O$. Each task $i$ has a given priority $p_i$, as well as known requirements for working capacity $b_{is}$ (referred to as bandwidth) depending on 
the shift $s$ it is assigned to. We note that we do not explicitly model operators but shifts of operators where the bandwidth requirement depends on the operator $o$ the shift $s$ belongs to. The set $S_o$ contains the shifts of operator $o$. Each task requires a specific skill level. Assignments to shifts of lower skilled operators incur a penalty of $d_{is}$. Operators which do not dispose of a certain skill at all cannot be assigned tasks requiring the respective skill. Each task belongs to one task group $\sigma$ out of the set of task groups $P$. The set of all tasks $V$ is composed of a set of mandatory tasks $U$ that have to be assigned and a set of optional tasks $W$  ($V = U \cup W$). 

We consider time to be discrete with an available number of time buckets $|T|$. 
Each operator defines a set of availability periods (referred to as availabilities). Within each availability at most one shift can be planned.
As such, each shift is limited by its corresponding availability given by earliest starting time $e_s$ and latest ending time $l_s$;  $T_s$ provides the set of time buckets in which shift $s$ can be planned. Likewise, each task $i$ has a set of time buckets $R_i$ it occupies. Then, the set $V_{st}$ gives the set of tasks that can be assigned to shift $s$ in time bucket $t$. 
In addition, a minimum shift length $\alpha_s$ and a maximum shift length $\omega_s$ is specified.  Every shift has requirements specific to each operator $o$ for break length $\gamma_o$ as soon as the maximum duration of a shift without break $\chi_s$ is exceeded, as well as rest length $\delta_o$ in between shifts of operator $o$. This implies that there may be availabilities in which no shifts are planned. The total break length may be consumed over multiple smaller breaks. However, each partial break has to have a minimum length $\beta_o$, depending on the operator $o$, and each separate break requires a preparation time $\rho$, during which an operator cannot perform any tasks. This preparation time does not count towards the length of the break. Each shift has further limitations on breaks by only allowing breaks after $\nu_s$ hours and only up until $\lambda_s$ hours into the shift. 
 The complete list of notation used for parameters and variables is given in Table \ref{tbl:params}. Using this notation, 
 we formulate the PTSP-TS as the following binary linear program:

\begin{table}[t!b]
\caption{Notation \label{tbl:params}}
\centering
\small
\begin{tabular}{lll}
\toprule
Sets and parameters && \\
\midrule
    $T$ & set of time buckets \\
    $U$ & set of mandatory tasks \\
    $W$ & set of optional tasks \\
	$V = U \cup W$ & set of all tasks \\
	$O$ & set of operators \\
	$S$ & set of shifts \\
    $S_o$ & set of shifts of operator $o$ \\
    $S_i$ & set of shifts that can cover task $i$ \\
    $T_s$ & set of time buckets of availability period in which shift
    $s$ can be planned\\
    $R_i$ & set of time buckets of task $i$ \\
    $V_{st}$ & set of tasks that can be assigned to shift $s$ in time bucket $t$ \\
	$P$ & set of task groups \\
	 \\
	$[e_s,l_s]$ & earliest starting time and latest ending time of shift $s$ (availability)\\
	$b_{is}$ & required bandwidth of task $i$ in shift $s$ \\
	$p_i$ & priority score of task $i$ \\
	$d_{is}$ & skill deviation of task $i$ if assigned to shift $s$ \\
	$\alpha_s$ & minimum duration of shift $s$ \\
	$\omega_s$ & maximum duration of shift $s$ \\
	$\chi_s$ & maximum duration of shift $s$ without requiring a break \\
	$\rho$ & preparation time \\
	$\beta_o$ & minimum partial break duration of operator $o$ \\
	$\gamma_o$ & minimum total break duration of operator $o$ \\
	$\delta_o$ & minimum rest duration of operator $o$\\
	$[\nu_s,\lambda_s]$ & time window of break for shift $s$\\
	$w_1$,$w_2$ & weights of objectives in the objective function \\
        $k$ & upper limit on the number of different groups for parallel tasks \\
\toprule
Decision variables && \\
\midrule
    \multicolumn{2}{l}{$x_{is}$=
	$\begin{cases}
		1, & \text{if task $i$ is  assigned to shift $s$}\  \\
		0, & \text{otherwise}
	\end{cases}$}\\
	\multicolumn{2}{l}{$u_{st}$=
	$\begin{cases}
		1, & \text{if shift $s$ is activated in time bucket $t$}\  \\
		0, & \text{otherwise}
	\end{cases}$} \\
	\multicolumn{2}{l}{$r_{st}$=
	$\begin{cases}
		1, & \text{if  shift $s$ deactivated in time bucket $t$}\  \\
		0, & \text{otherwise}
	\end{cases}$} \\
	\multicolumn{2}{l}{$v_{st}$=
	$\begin{cases}
		1, & \text{if time bucket $t$ in shift $s$ counts as break}\  \\
		0, & \text{otherwise}
	\end{cases}$} \\
	\multicolumn{2}{l}{$a_{ts\sigma}$=
	$\begin{cases}
		1, & \text{if task of group $\sigma$ is assigned in time bucket $t$ in shift $s$}\  \\
		0, & \text{otherwise}
	\end{cases}$} \\
	\multicolumn{2}{l}{$z_s$=
	$\begin{cases}
		1, & \text{if shift $s$ is enabled}\  \\
		0, & \text{otherwise}
	\end{cases}$} \\
	\multicolumn{2}{l}{$\hat{z}_s$=
	$\begin{cases}
		1, & \text{if shift $s$ requires a break}  \\
		0, & \text{otherwise}
	\end{cases}$} \\
\bottomrule
\end{tabular}
\end{table}

\allowdisplaybreaks

\begin{align}
    \max \hspace{2mm} w_1 \sum_{i\in V} \sum_{s \in S_i} p_i x_{is} - w_2 \sum_{i \in V} \sum_{s \in S_i} d_{is} x_{is} \label{objf}
\end{align}
subject to:
\begin{align}
	\sum_{s \in S_i} x_{is} & = 1 && \forall i \in U \label{eq1}\\
	\sum_{s\in S_i} x_{is} & \leq 1 && \forall i \in W \label{eq2}\\
	x_{is} & \leq z_s	&& \forall i \in V, s \in S_i \label{eq3}\\
	\sum_{i \in V_{st}} x_{is} b_{is} & \leq u_{st} && \forall s \in S, t \in T_s \label{eq4}\\
	u_{st+1} & \geq u_{st} && \forall s \in S, t \in T_s, t \leq |T_s|-1 \label{eq5}\\
	\sum_{t=e_s}^{l_s} u_{st}& \leq z_sM  && \forall s \in S \label{eq6}\\
	r_{st+1}& \geq r_{st} &&  \forall s \in S, t \in T_s, t \leq |T_s|-1\label{eq7}\\
	\sum_{i \in V_{st}} x_{is} b_{is} & \leq 1-r_{st} &&  \forall s \in S, t \in T_s \label{eq8}\\
	\sum_{t=e_s}^{l_s} r_{st} & \leq z_sM && \forall s \in S \label{eq9}\\
	\sum_{t=e_s}^{l_s} (u_{st} - r_{st}) & \geq \alpha_s z_s &&  \forall s \in S\label{eq10}\\
	\sum_{t=e_s}^{l_s} (u_{st} - r_{st}) & \leq \chi_s z_s + (\omega_s - \chi_s) \hat{z}_s &&  \forall s \in S\label{eq10b}\\
    \sum_{t'=t+1}^{t+\rho} \sum_{i \in V_{st'}} x_{is}b_{is} & \leq \rho - \rho v_{st} && \forall s \in S, t \in T_s, t \leq l_s-\rho \label{eq11}\\
    \sum_{i \in V_{st}} x_{is} b_{is} & \leq 1 - v_{st} &&  \forall s \in S, t \in T_s \label{eq12}\\
	\sum_{t=e_s}^{l_s} v_{st} & \geq \gamma_o\hat{z}_s && \forall o \in O, s \in S_o \label{eq13} \\
	v_{st} & \leq u_{st} - r_{st} && \forall s \in S, t \in T_s \label{eq14} \\
	v_{st} & \leq u_{s,t-\nu_s} && \forall s \in S, t \in T_s, t \geq \nu_s \label{eq15}\\
	v_{st} & \leq 1-u_{s,t-\lambda_s+1} && \forall s \in S, t \in T_s, t \geq \lambda_s \label{eq16}\\
	\sum_{t' = t+1}^{t + \beta_o -1} v_{st'} &\leq \beta_o (v_{st} + v_{s,t+\beta_o}) && \forall o \in O, s \in S_o, t \in T_s, t \leq |T_s|-\beta_o  \label{eq17}\\
	a_{ts\sigma} & \geq x_{is}  && \forall s \in S, i \in V, t \in R_i, \sigma \in P, \label{eq18}\\
	\sum_{\sigma \in P} a_{ts\sigma} & \leq k && \forall s \in S, t \in T_s \label{eq19}\\
	\sum_{t'=t+1}^{t+\delta_o} u_{s't'} & \leq (1-u_{st}+r_{st})\delta_o &&  \forall o \in O,  s \in S_o, s' \in {S_o|s' \neq s}, t \in T_s \label{eq20}\\
	x_{is} & \in \{0,1\}  &&\forall i \in V, S \in S_i \label{eq23}\\
	u_{st} & \in \{0,1\}  &&\forall s \in S, t \in T_s\\
	r_{st} & \in \{0,1\}  &&\forall s \in S, t \in T_s\\
	v_{st} &\in \{0,1\}  &&\forall s \in S, t \in T_s\\
	a_{ts\sigma} &\in \{0,1\}  && \forall s \in S, t \in T_s, \sigma \in P \\
	z_s & \in \{0,1\} &&\forall s \in S \\
	\hat{z}_s & \in \{0,1\} &&\forall s \in S \label{eq29}
\end{align}

The objective function \eqref{objf} consists of two parts: the sum of all assigned tasks 
weighted by their corresponding priority $p_i$ and the sum of all penalties that occur due to the assignment of tasks to operators with a lower than optimal skill level for the corresponding task group $d_{is}$. The overall function maximizes the difference between the total priority and the total penalty, each of which are individually weighted using $\mathnormal{w_1}$ and $\mathnormal{w_2}$. 

The first groups of constraints \eqref{eq1}, \eqref{eq2}, and \eqref{eq3} ensure the limits for the assignments of tasks. Tasks are split into mandatory tasks $U$ and optional tasks $W$. All mandatory tasks have to be assigned to exactly one single shift, while for all other tasks assignment is optional. Furthermore, constraints \eqref{eq3} require that a task may only be assigned to a shift if that shift has been enabled using the binary variable $z_s$.

Constraints \eqref{eq4}, \eqref{eq5}, and \eqref{eq6} set the limits for the binary variable $u_{st}$ depicting whether a given shift $s$ is active during the time bucket $t$. In constraints \eqref{eq4} the limit for the bandwidth usage is given. Each task requires a certain amount of an operator's attention referred to as bandwidth $b_{is}$. This bandwidth depends on the operator's skill as well as the task's complexity. The bandwidth used at any given time may only exceed zero if the shift is indeed active. Furthermore, as time is considered discrete, once a shift has been activated, the binary variables associated with the following time buckets will be $1$ until the final time bucket. Finally, constraints \eqref{eq6} ensure that the whole shift needs to be enabled using $z_s$ in order to be activated. If there is a surplus in operator capacity, an operator's shift can stay disabled, meaning he or she will not be working that shift.

Similar to the last four constraints, the shift deactivator $r_{st}$ has to be constrained. Constraints \eqref{eq7} ensure that once a shift $s$ has been deactivated, it will stay deactivated. In constraints \eqref{eq8} the limit for used bandwidth is set to zero once the shift $s$ has been deactivated. Constraints \eqref{eq9} state that a shift $s$ may only be deactivated if it is actually enabled.
Constraints \eqref{eq10} and \eqref{eq10b} ensure that the required minimum and maximum lengths per shift $(\alpha_s$ and $\omega_s)$ are respected and break requirements are met.

The following seven constraints ensure that each shift $s$ includes the required break times. Constraints \eqref{eq11} ensure that a preparation time for booting up to work after any break is considered. For an operator $o$ a time bucket $t$ can only be counted towards the required break time $\gamma_o$ if there are no assignments made after it for the length of the preparation time $\rho$. Constraints \eqref{eq12} ensure that the time bucket $t$ may only be counted as a break as long as the bandwidth is zero. Constraints \eqref{eq13} require that the sum of the break times has to be at least the minimum length if the shift length exceeds a certain limit. The next three sets of constraints $($\eqref{eq14} through \eqref{eq16}$)$ set the limits for whenever a time bucket may actually be counted as break. Constraints \eqref{eq14} require that only if a shift $s$ has been turned on, but not yet off, a break may be counted. Constraints \eqref{eq15} and \eqref{eq16} require all breaks to consider the earliest break starting time $\nu_s$ and the latest break end $\lambda_s$. Finally, constraints \eqref{eq17} require that for each operator $o$ any minimum break length requirements $\beta_o$ are respected.

Furthermore, the number of allowed task groups $P$ covered in a shift $s$ in a time bucket $t$ is restricted to $k$. Constraints \eqref{eq18} require that for each task group $\sigma$ the task group limit $a_{ts\sigma}$ is set to one if at least one task of that group is assigned at the time. Constraints \eqref{eq19} ensure that the total number of task group counting variables set to one at any time does not exceed the limit of $k$.

Between shifts covered by the same operator $o$, a rest time $\delta_o$ has to be maintained for legal reasons. Constraints \eqref{eq20} ensure that this rest time is respected in-between all shifts $S_o$ for each operator $o$. 

The domains of the variables are defined in \eqref{eq23}-\eqref{eq29}.\\

\section{Adaptive Large Neighborhood Search}
\label{method}
In order to solve the PTSP-TS, an adaptive large neighborhood search type algorithm is designed. Large neighborhood search (LNS) is a metaheuristic search concept proposed by \cite{shaw1998} and it is related to the ``ruin and recreate'' principle by \cite{schrimpf2000record}. It was then further developed for routing and scheduling problems by \cite{RP06a} with the introduction of an adaptive layer which changes the selection probability of  repair and destroy operators based on their past performance. 
Since then it has become one of the most popular heuristic search schemes in the routing and scheduling field \citep{pisinger2010large}. 

ALNS works as follows (see Algorithm \ref{alg:lns-allg}). In a first step, a feasible solution $s$ is generated. In every iteration, a removal operator $d$ and a repair operator $r$ are chosen based on their weights that change due to performance. In a next step, tasks are removed from $s$ using the chosen removal operator $d$ and put into the pool of not assigned tasks. Then, using the chosen repair operator $r$, tasks are (re-)inserted into the shift plan. In the case where the resulting solution $s$ meets the acceptance criteria, it replaces $s$. In the case where it is better than the best solution found so far, it replaces $s_{best}$. This is repeated until some stopping criterion is met.

\begin{algorithm}[H]
	\caption{ALNS \label{alg:lns-allg}}
	\begin{algorithmic}[1]
		\STATE generate a starting solution $s$;
		\STATE set $s_{best} := s$
		\REPEAT 
		\STATE choose a removal operator $d$ and a repair $r$ operator 
		\STATE  apply $d$  to  $s$ yielding $s'$
		\STATE  apply  $r$ to $s'$ yielding $s''$
		\STATE  decide if $s''$ is accepted as new incumbent solution; if yes $s \leftarrow s''$ 
		\STATE  check if $s''$ is better than $s_{best}$; if yes, $s_{best} \leftarrow s''$
		\STATE update the scores and weights of the operators
		\UNTIL a given stopping criterion is met
		\STATE return $s_{best}$
	\end{algorithmic}
\end{algorithm}

In the following we describe the different ingredients in further detail: the deployed destroy and repair heuristics, the developed randomization approach, the acceptance scheme as well as the adaptive mechanism.

\subsection{Destroy Heuristics}
We employ four different removal heuristics. .
Three of these heuristics - random, worst, and operator removal - are based on existing work \citep[e.g.,][]{KPDH10,RP06a} and tailored to the characteristics of 
the PTSP-TS.
The fourth operator, named time-based removal, exploits the fact that in our problem, tasks have fixed starting and ending times.  

\subsubsection{Random Removal}
\label{sec:randRem}
Random removal randomly chooses assigned tasks, removes them from the solution and adds them to the pool of unassigned tasks. 
It is implemented as follows. In a first step, a random number $l$ is drawn from the range \tt{chancelimits}. Then, for every currently assigned task, a random number $r$ is drawn and if  $r \leq l$, the respective task is removed from the schedule.

\subsubsection{Worst Removal}
Worst removal completely removes the shifts that contribute the lowest to the objective. It is specifically designed to select shifts that contribute unnecessarily high penalties, as well as shifts that are low on assignment density. The number of shifts destroyed is chosen randomly as follows: a random number $l$ is drawn from \tt{chancelimits} and let $N$ denote the number of currently planned shifts, then the worst $N l$ shifts are removed from the solution. 

\subsubsection{Operator Removal}
The operator removal algorithm removes all shifts of a given operator. It follows a similar logic as the random removal heuristic. Instead of removing individual tasks from the solution, a random number $r$ is drawn for each operator and in the case where $r \leq l$ (see Section \ref{sec:randRem}), all shifts of the respective operator are removed from the solution.

\subsubsection{Time-based Removal}
The time-based removal heuristic makes a vertical slice through the entire solution. The idea is to enable the switch of tasks between different shifts that would fit similar time slots. This removal method randomly selects a start time $t \in T$ as well as a length $l$ in the range \tt{widthlimits}, removing all assigned tasks $i$ for which $R_i \cap \{t, ..., t+l\} \neq \emptyset$, i.e., even if the task falls only partially into the time interval, it is removed. 

\subsection{Repair Heuristics}
All repair heuristics rely on sorting all possible task to operator assignments according to different criteria and then performing all feasible task to operator assignments in this order. The first three heuristics (A,B,C) rely on penalty and priority based sorting, the next heuristic (D) takes into account the objective function, and finally, the last heuristic (E) considers the value of assigning a task to an operator based on the time and bandwidth it will occupy. The heuristics are described in further detail in the following.

\subsubsection{Penalty and Priority Sorting}

The first three repair heuristics rely on sorting all task to operator assignments according to task $Priority$ $p_i$ and assignment $Penalty$ $d_{is}$. We ensure deterministic sorting of assignments by using a number of sequential sorting criteria. We use $Penalty$ as the main sorting factor in order to maximize the number of assignments to operators that have the desired skill level. $Priority$ is used as the second sorting value, with assignments with higher priority being preferred. The first repair heuristics both start sorting according to $Penalty$ ahead of $Priority$ and introduce the following additional tie breakers:

\begin{itemize}
	\item $A$: \{$Penalty,Priority,skillFitting$\}
	\item $B$: \{$Penalty,Priority,skillFitting*(-1)$\}
	\item $C$: \{$Penalty,Priority,Bandwidth*(-1),skillFitting$\}
\end{itemize}

As the first additional tie breaker we use a value called $skillFitting$. The calculation of this value differs depending on whether the operator fulfills the desired skill level of the considered task. If they do, we calculate the value of $skillFitting$ to be equal to $DesiredSkill - ActualSkill$. This gives preference to assigning tasks to operators that fulfill the skill requirement as closely as possible, in order to make sure  that the capacities of our highest-skilled operators are kept for the most difficult tasks. If the operator's skill level is lower than the desired level, then this value is equal to the $Penalty$ $d_{is}$. For our repair operators $A$ and $B$ we take $skillFitting$ as the third tie breaker, however for operator $B$ the value is inverted. This usage of the tie breaker has different objectives: While regular usage of the $skillFitting$ value tries to retain as much capacity of higher-skilled operators as possible, using the inverted value tries to fill up the shifts of the higher-skilled operators by using the tasks with the highest $Priority$. For sorting operator $C$ we introduce another differentiation value $Bandwidth$. $Bandwidth$ is the capacity a task occupies when assigned to a specific shift (denoted as $b_{is}$ above). As our problem allows for the assignment of multiple tasks in parallel, this value ranges from 1 (full capacity) down to $\frac{1}{6}$. The value of $Bandwidth$ for a specific assignment is dependent on the task as well as the operator's skill for parallel coverage. We use the inverted value as assignments using a smaller amount of capacity are generally preferred.

\subsubsection{Objective Sorting}
Sorting operator $D$ now combines the usage of $Penalty$ and $Priority$ into the value $Objective=w_1*p_i - w_2*d_{is}$ which is the actual value an assignment contributes to the objective function.

\begin{itemize}
\item $D$: \{$Objective,Bandwidth,skillFitting$\}
\end{itemize}

Similar to sorting operator $C$ we use $Bandwidth$ and $skillFitting$ as subsequent tie breakers.

\subsubsection{Efficiency Sorting}\label{eff}
\allowdisplaybreaks
For our fifth and final sorting operator $E$ we introduce a new value called $Efficiency$. This value is calculated using the following formula: $Efficiency = Objective$ $/(Bandwidth*Length)$. $Efficiency$ therefore gives us the contribution per occupied capacity of an operator. The $Efficiency$ of an assignment of tasks with the same priority grows the lower the occupied $Bandwidth$ of an operator and the shorter the task in general.

\begin{itemize}
	\item $E$: \{$Efficiency,Priority,Bandwidth*(-1),Groups*(-1)$\}
\end{itemize}

As first and second tie breaker we use the previously introduced values for $Priority$ and $Bandwidth$. However, due to the structure of the $Efficiency$ value, we require a tie breaker that is different from $skillFitting$. Here, we give preference to operators that have the least flexibility of covering different task groups. The value $Groups$ chosen as tie breaker is defined as the number of different task groups that an operator is skilled in. The skill levels are not considered here. As we aim at prioritizing assigning staff with low flexibility, the value of $Groups$ is inverted.

\subsection{Starting Solution}
To obtain a starting solution, we employ the repair operator Efficiency Sorting (described in \ref{eff}) to sort all tasks of the instance. We then construct an assignment by "repairing" the initially empty assignment, inserting tasks in the order provided by the sorting mechanism. Only feasible task assignments are considered.

\subsection{Acceptance Criteria}
For our implementation of ALNS, we consider three possibilities of handling a newly created solution. New solutions achieving a globally best objective value are immediately accepted and stored for further use. If a new solution has an objective value that is at maximum .5\% lower than the current incumbent solution, then the solution is accepted as new incumbent solution for the next iteration. All solutions more than .5\% worse than the current solution are discarded. In this case, the next iteration starts using the assignments of the incumbent solution. 

\subsection{Choice of Destroy and Repair Heuristics}
In accordance with \cite{RP06a}, we use separate weights $\eta_{h}$ for every possible repair and destroy heuristic $h$. Ahead of each iteration of the ALNS, destroy and repair heuristics are randomly chosen based on their weights. The result of each iteration (overall improvement, temporary improvement and rejection) is stored for the chosen heuristics. After a given amount of iterations $SegLen$, referred to as segment, the weights of all heuristics are updated based on their performance in the past segment. For the following segment, these updated values are used for randomized selection. The weights are updated periodically during the search, with the length of each segment staying the same.
The probability $\phi_{h}$ for a repair heuristic to be chosen for an iteration is calculated as follows:

\begin{align}
	\phi_{h} = \frac{\eta_{h}}{\sum_{{h}'=1}^{n_{h}}\eta_{{h}'}} 
\end{align}
The choice of destroy heuristic is made in the same way.

For updating the weights of the heuristics after each segment we use an approach that is based on \cite{RP06a}. Every repair and destroy heuristic receives an initial score $\sigma$ that increases in each iteration a heuristic was used, based on its performance. Scores are updated as follows:

\begin{align}
	\sigma = &
 \begin{cases} 
	\sigma + \epsilon_1, &\text{if $s''$ improves the global best solution } s_{best}, \\
	\sigma + \epsilon_2, &\text{if $s''$ improves the incumbent solution } s, \\
	\sigma + \epsilon_3, &\text{if $s''$ is accepted as new incumbent $s$ although it is worse},\\
	\sigma, &\text{otherwise. } \notag
\end{cases}
\end{align}

In our computational study, we evaluate this approach, using different parameter settings and we also consider keeping $\epsilon_3 = 0$, which results in no reward for solutions that are worse than the current incumbent, even if they were accepted. 
Each segment consists of a total of 100 iterations. At the end of each segment, the weight of each destroy and repair heuristic is updated using the following formula:

\begin{align}
	\eta_h = (1-p_{react})\eta_{h}+p_{react}\frac{\sigma_{h}}{\max(1,\Omega_{h})}
\end{align} 

where $\Omega_{h}$ denotes the number of times the considered repair heuristic was used and $p_{react}$ is a predetermined weight controlling the influence of the new score on the historic weight.

\subsection{Task Stacking}
To improve bandwidth usage, we introduce task stacking into our repair heuristics. 
$Stacking$ causes the repair algorithm to try to build stacks of similar tasks (the seed tasks being the one that is supposed to be inserted) and to insert all stacked tasks at once. Tasks are considered for stacking if they share starting and ending times and belong to the same task group as the seed task. 
Since stackable tasks are assigned in parallel, in the best case, an operator's capacity is fully used by the chosen tasks.

\subsection{Slice Shuffling}
In order to introduce some randomization into our repair heuristics, we develop a slice shuffling mechanism. As all possible assignments are always sorted in the same way whenever a given repair heuristic is applied, we introduce a semi-random shuffling operator in order to prevent the algorithm from generating the same solutions over and over. Slice shuffling takes the sorted list of possible assignments and divides it into slices of length $\mu$.  
The order of the assignments within a given slices is then shuffled. Preliminary experiments with different parameter settings have led to the definition of the set $sl$. The set $sl$ contains all possible values $\mu$ may take for a given iteration: at the start of each repair step, the value for $\mu$ is randomly chosen out of $sl$. Making sure that the slices are not too large allows us to explore different options for assignments, while ensuring that the initial sorting done by the repair heuristic does not become pointless. Too small slices may produce very similar results due to the potentially large number of tasks that do not differ in their sorting value.

\section{Computational Results}
\label{results}
The ALNS algorithm as well as the MIP model were implemented in Python, with the MIP model using the CPLEX 12.8 as MIP-solver. All tests are conducted on a computing cluster with 64 nodes consisting of 2 Intel Xeon X5570 CPUs with 4 cores as well as 48 GB of RAM each. Each run uses a single thread. All ALNS results are based on ten random runs per instance and we use a run time limit of 6 hours for the MIP-solver.  
Each time-bucket in the MIP represents five minutes.

\subsection{Data Sets}
The test instances used in this paper are based on information from our company partner Sportradar as well as on real-world sports data. Sportradar's operators reside in four different countries giving rise to different working and break length regulations. The preparation time $\rho$ is assumed to be 20 minutes irrespective of the working location. Tasks are sports matches and belong to one out of 22 different task groups and may require a maximum skill level of 100. In terms of bandwidth, a task may require the complete bandwidth (1) of an operator or only a fraction of the bandwidth. The minimum is $\frac{1}{6}$. This implies that the maximum number of parallel tasks is six. Furthermore, at most tasks from two different task groups may be scheduled in parallel. 
The generated test instances are divided into four different groups with ten instances each. Small instances contain only a handful of workers with around 100 tasks to be chosen for assignment and the considered planning horizon is limited to 8 hours. Medium-sized instances contain up to ten workers, between 400 and 700 tasks and a total planning horizon of 18 hours.  
The set of large instances contains a higher number of operators and is set to reflect the assignments for an entire day with a considered planning horizon of 24 hours and between 600 and 1100 tasks. Finally, ten very large-scale instances include shifts and tasks for a time span of 72 hours, this results in data sets with between 180 and 230 shifts and between 2000 and 3300 tasks. The length of the planning horizon corresponds roughly to the requirement of Sportradar which schedules their shifts three days in advance.
Instance names are given in the following format: $|T|$\_$|S|$\_$|V|$ providing the planning horizon of the instance in hours ($|T|$), the number of possible shifts ($|S|$), and the number of available tasks ($|V|$).

\subsection{Results for Benchmark Instances}
In a first step, we evaluate the performance of the developed ALNS on those instances that can also be tackled by the MIP-solver. 

Based on preliminary tests, all ALNS parameters have been set. The set of parameters is \{\tt{chancelimits}, \tt{widthlimits}, $sl$, $\rho$, $\epsilon_1$, $\epsilon_2$, $\epsilon_3$, $p_{react}$\}. 
The range \tt{chancelimits} which is used to define the probability for a task to be removed from the solution in the random removal operator, and the probability of an operator's shifts to be deleted from the solution is set to $(0.1,0.5)$. The range \tt{widthlimits} from which the value of the slice length in the time based removal operator is drawn is set to [180,300] (minutes).
The set $sl$ contains all possible values for the length of the slice $\mu$ which is used in the slice shuffling step. This set is defined to be \{0,4,8,16\}. 
The parameters used to reward destroy and repair operators for good performance \{$\epsilon_1$, $\epsilon_2$, $\epsilon_3$\} are initially set to \{40,25,8\}.  However, in the following we also test the setting \{40,25,0\}, i.e., no reward for deteriorating solutions that are accepted as new incumbent solutions. The value of $p_{react}$ is set to 0.1 for all computations.
The base setting of the weights in the objective function is $w_1 = 1$ and $w_2 = 1$.

Table~\ref{tbl:mipcomp} reports average values over small and medium-sized instances and the average average gap (AGap) as well as the average best gap (BGap) with respect to the MIP solution for setting $\epsilon_3 = 8$ and setting $\epsilon_3 = 0$ and different ALNS iteration limits (Iterations).

\begin{table}[tbh]
\setlength{\tabcolsep}{2ex}
\caption{Comparison of average solutions obtained by the ALNS using different values for $\epsilon_3$ in comparison to the best solution values obtained by the MIP-solver for small and medium-sized instances.\label{tbl:mipcomp}}
\centering
\begin{tabular}{lrrrr}
\toprule
\multicolumn{5}{l}{ALNS}\\
\midrule
Setting & Iterations & AGap  & BGap & Avg \\
\midrule
$\epsilon_3 = 8$ & 1,000& 3.27 & 1.78 & 5123.4\\
 &10,000& 1.57 & 0.63  & 5226.2 \\
&25,000& 1.14 & 0.39  & 5252.5 \\
&100,000& 0.67 & 0.09  & 5286.1 \\
\midrule
$\epsilon_3 = 0$ & 1,000& 3.22 & 1.98 & 5127.2\\
 &10,000& 1.48 & 0.52  & 5231.7\\
&25,000& 1.19 & 0.34 & 5249.8\\
&100,000& 0.91 & 0.22 & 5269.2\\
\bottomrule
\end{tabular}
\end{table}

Detailed results on a per instance level, can be found in the appendix in Tables \ref{tablestart} to \ref{tableend}, where Tables \ref{tablestart} to \ref{firstend} contain the results for setting $\epsilon_3 = 8$ and Tables \ref{secondstart} to \ref{tableend} those of setting $\epsilon_3 = 0$. All tables are structured as follows:  
for each instance, we provide the average objective value ($Avg$), the best objective value ($Best$), the average run time ($Time$) in seconds as well as the average ($AGap$) and the best gap ($BGap$) of the objective value computed by the ALNS to the objective value obtained with the MIP-solver in percent. Furthermore, we provide the average number of tasks assigned ($|M|$). For the solutions obtained with the MIP-solver, we give the best obtained lower bound ($LB$) and the run time ($Time$) in seconds. For instances that the MIP-solver did not solve within the time limit of six hours, we report the optimality gap ($MIPGap$) given by CPLEX at the end of the run time. In the final column we provide the number of tasks assigned in the best solution obtained by the MIP-solver ($|M|$). For each set of benchmark instances, we give the results for 1,000, 10,000, 25,000 as well as 100,000 iterations in separate tables. In each table, we add a summary line that provides the average of the average gaps as well as the average of the best gaps achieved over the instances of the benchmark set.

As the solutions obtained by the MIP-solver (given in Table \ref{tablestart} and repeated in Tables \ref{table10Tits} to \ref{tableend} for comparison purposes) show, most problems of the smallest size with a planning horizon of eight hours can be solved to optimality within the time limit. However, even for medium-sized instances that consider a planning horizon of 18 hours, the MIP-solver already reaches its limits. For large instances considering a planning horizon of a full day, the MIP implementation already fails at the root node. Therefore, 
in this initial set of experiments, we only report results for small and medium-sized instances. 

Overall, setting $\epsilon_3=8$ 
performs slightly better than setting $\epsilon_3=0$. This difference 
mainly 
becomes visible for the highest number of iterations when comparing the best objective values found. For the highest number of iterations tested, the best solutions obtained using $\epsilon_3=8$ are within less than 0.1 percent of the solution obtained by the MIP-solver. However, we note that this average consists of solutions where the heuristic significantly outperforms the solution of the MIP as well as some solutions where it falls off slightly. The maximum percentage difference of the heuristic beating the MIP implementation and vice versa is at about 1.5 percent. We note that while the MIP-solver was given an upper limit of six hours to calculate the solution, we chose the limit of 100,000 iterations for our longest comparison to stay considerably below 30 minutes of run time for the heuristic. We also note that our implementation was made in Python 3. Speed-ups of a factor of 10 may be achieved when using a different programming language \citep{tricoire2021broute}.
In the following, all reported ALNS results use the setting $\epsilon_3 = 8$.

\subsection{Additional Objectives and Trade-off Analysis \label{addobj}}

In addition to the two components of the objective function considered in the PTSP-TS, Sportradar was interested in optimizing an additional three objectives introduced in detail in the following paragraphs. We name them $Groups$, $Consec$, and $Workload$.

Tasks can be grouped according to their similarity. When similar tasks are performed within the same shift, the quality of the service increases. Adversely, the more different tasks from different groups a worker is assigned during a shift, the lower the quality the worker will be able to provide. The third objective is therefore to minimize the number of different task groups assigned to a full shift and is named $Groups$. While there is an upper limit to the number of tasks of different groups for concurrent tasks, this objective seeks to minimize the number for the whole shift. 

The fourth objective introduces the idea of consecutive tasks $Consec$ and adds a bonus if they are scheduled in the same shift. As some groups of tasks may feature tasks that should start back-to-back, it is generally helpful to assign these consecutive tasks to the same shift. This information is not directly exploited in the base setting of our ALNS. We explain below, how this information can be incorporated into the ALNS, so as to increase the number of consecutive tasks assigned to the same shift.

The final objective is to minimize the amount of employee idleness by maximizing the workload in percent of total available time. We name this measure $Workload$. This objective was omitted in the model proposed in this paper, as it is generally considered to be optimized through the choice of the most valuable tasks. Introducing this objective might incur assignments of lower priority tasks in order to reduce employee idle time. 

In order to evaluate the impact of considering these additional objectives, we evaluate different weight settings. Let \{$Priority$, $Penalty$, $Groups$, $Consec$, $Workload$\} denote the set of considered objectives. We combine them into a weighted sum where each objective receives a different weight, depending on the chosen focus. Weight setting \tt{Control} is the benchmark setting. It uses the same weights as above, i.e., \{1,1,0,0,0\}. Weight setting \tt{Priority} focuses on the first objective only, using weight set \{1,0,0,0,0\}. Setting \tt{Penalty} puts a comparably high weight on the second objective, using weight set \{1,100,0,0,0\}. The weights for setting \tt{TaskGroups}, where emphasis is put on grouping similar tasks together, are \{1,1,-10,0,0\}. We note that the $Groups$ and $Penalty$ objectives are only considered as secondary objectives, as obviously without optimization of $Priority$, the value of the $Groups$ objective would be 1 for solutions without parallel tasks of different groups and 0 (the best possible value) for solutions without any tasks assigned. The latter also holds for the $Penalty$ objective. The employed weights for the maximization of the $Workload$ objective (setting \tt{Workload}) are \{1,1,0,0,10000\}. 

Finally, we analyze a setting where all five objectives considered by the company receive a weight, so as to roughly reflect their importance from the company perspective: $Priority$ is the main objective, $Penalty$ and $Workload$ are secondary objectives, while $Groups$ and $Consec$ are the least important objectives. In this experiment, we introduce the concept of successor matching into the proposed ALNS: we add the possibility upon assigning a match to assign its direct predecessor/successor at the same time and to the same shift (if feasible). This is an additional step taken to improve the number of consecutive matches that are assigned to the same operator. Furthermore, in the destruction step of the ALNS, we introduce successor destruction, which allows us to destroy any assigned successors and predecessors of a destroyed match, even if they are assigned to different shifts in the current solution. Successor destruction is randomly turned on or off at the beginning of a new destroy step. These final tests are made using weights for the five objectives of \{1,1,-1,1,1000\}. We denote the ALNS with these additional ingredients ALNS+ and the weight setting \tt{AllObjectives}.

\begin{table}[tbh]
\setlength{\tabcolsep}{1.2ex}
\caption{Comparison of average solutions across all instances, obtained with 100,000 ALNS iterations using different weight settings. For setting \tt{AllObjectives} ALNS+ is used. \label{tbl:comparison}}

\small
\centering
\begin{tabular}{lp{3pt}rrrrrrr}
\toprule
 && \multicolumn{5}{l}{Average individual objective values}\\
 \cmidrule{3-7}
Weight setting && $Priority$ & $Penalty$ & $Groups$ & $Consec$ & $Workload$ & $OF$ & $OFsmall$\\
\midrule
\tt{Control} && 44733.6 & -206.55 & 2.34 & 11.05 & 0.86 & 44527.05 & 5286.10\\
\midrule
\tt{Priority} && 44875.28 & -482.92 & 2.34 & 11.00 & 0.86 & 44392.35 & 5264.68\\
\tt{Penalty} && 43998.70 & -0.00 & 2.32 & 10.70 & 0.86 & 43998.70 & 5234.90\\
\tt{TaskGroups} && 44728.35 & -206.06 & 2.28 & 11.18 & 0.86 & 44522.29 & 5288.38\\
\tt{Workload} && 44576.25 & -195.04 & 2.38 & 10.89 & 0.88 & 44381.21 & 5168.35\\
\midrule
\tt{AllObjectives} && 44751.4 & -202.16 & 2.32 & 23.55 & 0.87 & 44549.24 & 5288.87\\
\bottomrule
\end{tabular}
\end{table}

We summarize the results of the different weight settings in Table \ref{tbl:comparison}. Average results per instance can be found in Tables \ref{tbl:priority}--\ref{tbl:allobj}  in the appendix.

In Table \ref{tbl:comparison}, in addition to the average values for every single objective, we report the average values obtained for the original objective function ($OF$) as well as the average value of the original objective function considering only the small and medium-sized instances ($OFsmall$). The values given in the table show, that each of the four specialized settings reaches the best value for the considered objective among all approaches tested. Furthermore, the \tt{Control} setting, taking into account only priority and penalty, provides the best average value of the original objective function, considering all test instances. When considering small and medium-sized instances only, additional weight on the minimization of task groups (setting \tt{Taskgroups}) has a positive side effect on the penalty objective.

Finally, setting \tt{AllObjectives} 
obtains slightly better solutions on average for the original objective function, which indicates that the additional ingredients, focusing on consecutive matches, help the ALNS in finding even slightly better solutions for the original objective function by planning more consecutive matches together.

The results obtained for the different weight settings allow us to derive some managerial insights and pointers towards further use and practical applications. First, they 
indicate that secondary objectives may be considered at a comparatively small price when regarding the two main objectives. All results considering focused objective functions share a significant improvement in the focused objective over the baseline or control setting. 

Second, additional ingredients focusing on the planning of consecutive matches together with weights on all objectives (setting \tt{AllObjectives}) has a positive impact. It returns solutions that improve all three secondary objectives 
and even slightly improve the values on the two main objectives on average. This implies that additional information (in our case, the information relating to matches that can and should be planned consecutively) should be used and exploited in the design of the algorithm. 

In the following case study, where the proposed algorithm is tested on company data, ALNS+, as deployed in the \tt{AllObjectives} setting, is used, and compared to ALNS with the setting \tt{Control}.

\section{Case Study}
\label{study}

The considered problem was brought to us by Sportradar. Operators are to be assigned sports matches to supervise and determine whether provided betting odds reflect reality. The considered problem includes all five of the previously presented objectives. Furthermore, Sportradar requires a time horizon of three days to be planned at once. The comparison made in this section is only performed between the solutions obtained with ALNS+ with the setting \tt{AllObjectives}, ALNS with the setting \tt{Control} and the assignments the company has made.

\begin{table}[tbh]
\setlength{\tabcolsep}{2ex}
\caption{Comparison of original company assignments with average solutions obtained with ALNS+ using weight setting \tt{AllObjectives} on case study data,
as well as weighting the two main objectives only (\tt{Control}) using ALNS \label{tbl:rw}}
\centering
\begin{tabular}{lp{3pt}rrr}
\toprule
Objective && Company & \tt{AllObjectives} &
\tt{Control}\\
\midrule
Total Priority ($Priority$) && 111390 & 179902.0 & 179865.0\\
\\
Total Penalty ($Penalty$) && -10577 & -752.6 & -816.7\\
\\
Average sports ($Groups$) && 1.982 & 2.553 & 2.515\\
Maximum sports && 5 & 6.2 & 6.1\\
\\
Consecutive same ($Consec$) && 12 & 23.2 & 18.7\\
Consecutive different && 16 & 82.4 & 82.9\\
Consecutive scheduled && 22 & 82.6 & 83.0\\
Consecutive available && 83 & 83.0 & 83.0\\
\\
Ratio workingtime/total ($Workload$)&& 0.822 & 0.841 & 0.841\\
\\
Scheduled matches && 1584 & 2684.8 & 2696.6\\
Total matches && 5266 & 5266.0 & 5266.0\\
\\
Objective Value ($OF$) && 100813 & 179149.4 & 179048.3\\
\bottomrule
\end{tabular}
\end{table}

As the basis for the comparison between the performance of our solution method and the original company assignments, we use company data spanning a total of five days. The resulting data set comprises a total of 6706 matches assignable to a total of 379 potential shifts which belong to 124 operators. To increase the comparability between test assignments and original assignments, as well as to simulate the continuity of the real world problem, we choose to exclude assignments made for the first and the final of these five days from contributing to the objective function. This leaves a total of 5266 matches available for assignment within the time frame of three days. Table \ref{tbl:rw} provides this comparison on real-world data using the proposed ALNS+ algorithm, the ALNS algorithm and the assignments made by the company.  
ALNS is run 
only considering the two main objectives (weight setting \tt{Control}) while 
ALNS+ uses weights for all five objectives (weight setting \tt{AllObjectives}). The values provided in the table are obtained for an iteration limit of 100,000 iterations for both  weight settings. 
The comparison shows that assignments made using the proposed method significantly outperform past company assignments when regarding the two main objectives. The third objective $Groups$ shows that our ALNS+ assigns more tasks of different task groups than the company solution, which is only treated as a secondary objective. The same holds true for the highest value ("Maximum Sports") over all shifts. Comparison for the objective of assigning consecutive matches $Consec$ shows that the generated assignments include many more consecutive matches than the company solution. In addition to the actual objective of consecutive matches assigned to the same shift, we further report the values for consecutive matches assigned to any shift ("Consecutive different"), including same and different shifts, as well as consecutive matches where at least one of the two matches is assigned in the overall solution ("Consecutive scheduled"). The reported ratio of working time to total time ($Workload$) shows that our assignments achieve an average of 0.841 for both settings, which is significantly higher than the value of the company assignments. Finally, we report the number of scheduled matches in the solutions, as well as the value obtained for the objective function ($OF$) used in the \tt{Control} setting.

Summarizing the above, the proposed heuristic significantly outperforms a comparable assignment made by Sportradar in all considered performance measures but one ($Groups$). The large gap in the objective value of the main two objectives can partially be explained by the limit of the scale of importance of each of the considered matches. As the matches are generally assigned a priority between 0 and 100, the company solution considered a direct approach to this goal. A task with a higher priority was therefore assigned as early as possible. However, long matches with a high value for priority may be inferior to short matches with an average value due to the amount of working capacity they require from a given worker. This is very clearly displayed in the comparison when regarding the values for the fifth objective, minimizing worker idle time and the total number of scheduled matches in the solutions. While our solutions assign more than 1,000 additional matches, the ratio of working time to total time on average is only around two percent higher than in the company solution.  

\section{Conclusions}
\label{conc}
In this paper we have modeled and solved a personnel task scheduling problem with task selection (PTSP-TS). We have developed a tailored adaptive large neighborhood search algorithm. Repair and destroy operators previously used in the literature have been adapted to fit the proposed problem. Additionally, we propose new and tailored operators. Furthermore, we have proposed a mathematical model of the considered problem and benchmark the ALNS algorithm against solving the MIP formulation with CPLEX. The required run time of CPLEX is long even for medium-sized instances and cannot be used for time horizons interesting for real-world application. The proposed ALNS has proven very competitive compared to the previous planning at Sportradar
and has since been implemented for regular use.\\
Future research will focus on further analyzing the trade-off between the different objectives, e.g., by considering a subset of the objectives concurrently in a multi-objective approach which can produce (an approximation of) the Pareto frontier.

\section*{Acknowledgements}
This research was funded in whole, or in part, by the Austrian Science Fund (FWF) [P 31366]. For the purpose of open access, the author has applied a CC BY public copyright licence to any Author Accepted Manuscript version arising from this submission.
We also wish to thank Sportradar for having provided us with the real-world data.

\bibliographystyle{apalike}      
\bibliography{library}  

\newpage
\appendix
\section{Detailed Computational Results}
\vspace{-4ex}
\setlength{\tabcolsep}{1ex}
\begin{table}[t!bh]
\caption{Results for small and medium-sized instances (1,000 ALNS iterations, $\epsilon_3 = 8$)\label{tablestart}}
\scriptsize
\centering
\begin{tabular}{lp{3pt}rrrrrrp{3pt}rrrr}
\toprule
&&\multicolumn{6}{c}{LNS}&&\multicolumn{4}{c}{MIP}\\ \cmidrule{3-8}\cmidrule{10-13}
Instance && Avg & Best & Time & AGap & BGap & $|M|$ && MIP & Time & MIPGap & $|M|$\\
\midrule
8\_5\_101 && 3400.6 & 3469.0 & 7.1 & 3.09 & 1.14 & 52.6 && 3509 & 942 & 0 & 44\\
8\_4\_160 && 4176.8 & 4229.0 & 8.2 & 2.09 & 0.87 & 60.6 && 4266 & 459 & 0 & 53\\
8\_2\_119 && 1449.0 & 1460.0 & 4.3 & 0.75 & 0.0 & 22.0 && 1460 & 0.5 & 0 & 22\\
8\_5\_72 && 2681.7 & 2704.0 & 6.0 & 1.91 & 1.1 & 35.0 && 2734 & 2 & 0 & 35\\
8\_2\_109 && 2307.0 & 2360.0 & 4.9 & 2.25 & 0.0 & 29.0 && 2360 & 14 & 0 & 30\\
8\_4\_172 && 4968.0 & 5060.0 & 7.9 & 5.37 & 3.62 & 57.0 && 5250 & 753 & 0 & 61\\
8\_5\_124 && 3430.0 & 3565.0 & 6.5 & 3.79 & 0.0 & 47.3 && 3565 & 82 & 0 & 40\\
8\_5\_145 && 4610.6 & 4656.0 & 8.0 & 2.32 & 1.36 & 66.4 && 4720 & 21600 & 3.97 & 69\\
8\_5\_80 && 2619.6 & 2689.0 & 6.3 & 3.62 & 1.07 & 33.1 && 2718 & 61 & 0 & 33\\
8\_6\_136 && 4234.5 & 4285.0 & 8.7 & 2.32 & 1.15 & 75.2 && 4335 & 21600 & 0.62 & 75\\
18\_10\_567 && 8970.2 & 9135.0 & 18.2 & 6.41 & 4.69 & 127.9 && 9585 & 21600 & 6.17 & 135\\
18\_9\_548 && 9775.0 & 9926.0 & 16.0 & 4.41 & 2.93 & 137.9 && 10226 & 21600 & 8.51 & 141\\
18\_4\_412 && 3344.0 & 3420.0 & 8.1 & 2.79 & 0.58 & 37.6 && 3440 & 12560 & 0 & 39\\
18\_6\_440 && 5945.6 & 6007.0 & 9.7 & 3.12 & 2.12 & 68.9 && 6137 & 21600 & 4.31 & 70\\
18\_4\_404 && 4798.0 & 4865.0 & 8.9 & 3.56 & 2.21 & 63.7 && 4975 & 21600 & 4.32 & 66\\
18\_8\_514 && 9003.0 & 9178.0 & 13.0 & 4.04 & 2.17 & 102.6 && 9382 & 21600 & 4.12 & 108\\
18\_10\_710 && 8569.5 & 8755.0 & 18.7 & 5.2 & 3.15 & 116.5 && 9040 & 21600 & 10.55 & 119\\
18\_8\_557 && 8202.4 & 8400.0 & 14.4 & 5.78 & 3.51 & 110.8 && 8706 & 21600 & 8.66 & 116\\
18\_4\_363 && 3333.0 & 3337.0 & 7.3 & 0.12 & 0.0 & 39.8 && 3337 & 43 & 0 & 40\\
18\_9\_368 && 6627.9 & 6677.0 & 11.1 & 2.44 & 1.72 & 81.0 && 6794 & 21600 & 7.85 & 81\\
\midrule
Average &&&&& 3.27 & 1.67 &&&&&\\
\bottomrule
\end{tabular}
\end{table}
\begin{table}[t!bh]
\caption{Results for small and medium-sized instances (10,000 ALNS iterations, $\epsilon_3 = 8$)
\label{table10Tits}}
\scriptsize
\centering
\begin{tabular}{lp{3pt}rrrrrrp{3pt}rrrl}
\toprule
&&\multicolumn{6}{c}{LNS}&&\multicolumn{4}{c}{MIP}\\ \cmidrule{3-8}\cmidrule{10-13}
Instance && Avg & Best & Time & AGap & BGap & $|M|$ && MIP & Time & MIPGap & $|M|$\\
\midrule
8\_5\_101 && 3448.9 & 3479.0 & 30.6 & 1.71 & 0.85 & 52.8 && 3509 & 942 & 0 & 44\\
8\_4\_160 && 4211.3 & 4229.0 & 33.7 & 1.28 & 0.87 & 60.5 && 4266 & 459 & 0 & 53\\
8\_2\_119 && 1459.0 & 1460.0 & 14.7 & 0.07 & 0.0 & 22.0 && 1460 & 0.5 & 0 & 22\\
8\_5\_72 && 2700.5 & 2734.0 & 24.0 & 1.23 & 0.0 & 35.1 && 2734 & 2 & 0 & 35\\
8\_2\_109 && 2332.0 & 2360.0 & 17.1 & 1.19 & 0.0 & 29.2 && 2360 & 14 & 0 & 30\\
8\_4\_172 && 5092.0 & 5230.0 & 37.2 & 3.01 & 0.38 & 58.8 && 5250 & 753 & 0 & 61\\
8\_5\_124 && 3532.8 & 3565.0 & 27.2 & 0.9 & 0.0 & 48.0 && 3565 & 82 & 0 & 40\\
8\_5\_145 && 4665.8 & 4680.0 & 34.5 & 1.15 & 0.85 & 67.3 && 4720 & 21600 & 3.97 & 69\\
8\_5\_80 && 2669.3 & 2705.0 & 24.1 & 1.79 & 0.48 & 33.3 && 2718 & 61 & 0 & 33\\
8\_6\_136 && 4288.5 & 4315.0 & 38.4 & 1.07 & 0.46 & 75.6 && 4335 & 21600 & 0.62 & 75\\
18\_10\_567 && 9301.9 & 9485.0 & 117.7 & 2.95 & 1.04 & 131.3 && 9585 & 21600 & 6.17 & 135\\
18\_9\_548 && 10001.4 & 10056.0 & 101.2 & 2.2 & 1.66 & 140.0 && 10226 & 21600 & 8.51 & 141\\
18\_4\_412 && 3422.0 & 3440.0 & 39.9 & 0.52 & 0.0 & 38.4 && 3440 & 12560 & 0 & 39\\
18\_6\_440 && 6019.3 & 6037.0 & 52.9 & 1.92 & 1.63 & 69.8 && 6137 & 21600 & 4.31 & 70\\
18\_4\_404 && 4864.0 & 4960.0 & 47.1 & 2.23 & 0.3 & 64.6 && 4975 & 21600 & 4.32 & 66\\
18\_8\_514 && 9197.6 & 9316.0 & 78.6 & 1.97 & 0.7 & 104.8 && 9382 & 21600 & 4.12 & 108\\
18\_10\_710 && 8804.4 & 8924.0 & 127.3 & 2.61 & 1.28 & 118.2 && 9040 & 21600 & 10.55 & 119\\
18\_8\_557 && 8441.9 & 8510.0 & 91.0 & 3.03 & 2.25 & 113.5 && 8706 & 21600 & 8.66 & 116\\
18\_4\_363 && 3337.0 & 3337.0 & 36.1 & 0.0 & 0.0 & 40.0 && 3337 & 43 & 0 & 40\\
18\_9\_368 && 6741.7 & 6817.0 & 66.1 & 0.77 & -0.34 & 81.3 && 6794 & 21600 & 7.85 & 81\\
\midrule
Average &&&&& 1.58 & 0.62 &&&&&\\
\bottomrule
\end{tabular}
\end{table}
\begin{table}[tbh]
\caption{Results for small and medium-sized instances (25,000 ALNS iterations, $\epsilon_3 = 8$)}
\scriptsize
\centering
\begin{tabular}{lp{3pt}rrrrrrp{3pt}rrrl}
\toprule
&&\multicolumn{6}{c}{LNS}&&\multicolumn{4}{c}{MIP}\\ \cmidrule{3-8}\cmidrule{10-13}
Instance && Avg & Best & Time & AGap & BGap & $|M|$ && MIP & Time & MIPGap & $|M|$\\
\midrule
8\_5\_101 && 3466.0 & 3509.0 & 64.8 & 1.23 & 0.0 & 52.9 && 3509 & 942 & 0 & 44\\
8\_4\_160 && 4222.2 & 4229.0 & 70.6 & 1.03 & 0.87 & 60.2 && 4266 & 459 & 0 & 53\\
8\_2\_119 && 1460.0 & 1460.0 & 30.2 & 0.0 & 0.0 & 22.0 && 1460 & 0.5 & 0 & 22\\
8\_5\_72 && 2700.5 & 2734.0 & 49.3 & 1.23 & 0.0 & 35.1 && 2734 & 2 & 0 & 35\\
8\_2\_109 && 2337.0 & 2360.0 & 34.7 & 0.97 & 0.0 & 29.2 && 2360 & 14 & 0 & 30\\
8\_4\_172 && 5141.0 & 5250.0 & 81.8 & 2.08 & 0.0 & 59.3 && 5250 & 753 & 0 & 61\\
8\_5\_124 && 3558.0 & 3565.0 & 57.6 & 0.2 & 0.0 & 48.0 && 3565 & 82 & 0 & 40\\
8\_5\_145 && 4676.8 & 4680.0 & 71.8 & 0.92 & 0.85 & 67.2 && 4720 & 21600 & 3.97 & 69\\
8\_5\_80 && 2690.4 & 2706.0 & 49.2 & 1.02 & 0.44 & 33.2 && 2718 & 61 & 0 & 33\\
8\_6\_136 && 4303.5 & 4315.0 & 83.8 & 0.73 & 0.46 & 75.7 && 4335 & 21600 & 0.62 & 75\\
18\_10\_567 && 9337.9 & 9525.0 & 285.0 & 2.58 & 0.63 & 131.4 && 9585 & 21600 & 6.17 & 135\\
18\_9\_548 && 10104.8 & 10240.0 & 238.6 & 1.19 & -0.14 & 139.9 && 10226 & 21600 & 8.51 & 141\\
18\_4\_412 && 3422.0 & 3440.0 & 90.9 & 0.52 & 0.0 & 38.4 && 3440 & 12560 & 0 & 39\\
18\_6\_440 && 6029.3 & 6037.0 & 119.9 & 1.75 & 1.63 & 70.0 && 6137 & 21600 & 4.31 & 70\\
18\_4\_404 && 4891.5 & 4960.0 & 105.8 & 1.68 & 0.3 & 64.7 && 4975 & 21600 & 4.32 & 66\\
18\_8\_514 && 9262.4 & 9332.0 & 178.8 & 1.27 & 0.53 & 106.2 && 9382 & 21600 & 4.12 & 108\\
18\_10\_710 && 8884.8 & 9010.0 & 307.7 & 1.72 & 0.33 & 119.9 && 9040 & 21600 & 10.55 & 119\\
18\_8\_557 && 8499.0 & 8560.0 & 215.3 & 2.38 & 1.68 & 113.7 && 8706 & 21600 & 8.66 & 116\\
18\_4\_363 && 3337.0 & 3337.0 & 80.6 & 0.0 & 0.0 & 40.0 && 3337 & 43 & 0 & 40\\
18\_9\_368 && 6765.9 & 6819.0 & 155.1 & 0.41 & -0.37 & 81.8 && 6794 & 21600 & 7.85 & 81\\
\midrule
Average &&&&& 1.15 & 0.36 &&&&&\\
\bottomrule
\end{tabular}
\end{table}
\begin{table}[tbh]
\caption{Results for small and medium-sized instances (100,000 ALNS iterations, $\epsilon_3 = 8)$\label{firstend}}
\scriptsize
\centering
\begin{tabular}{lp{3pt}rrrrrrp{3pt}rrrl}
\toprule
&&\multicolumn{6}{c}{LNS}&&\multicolumn{4}{c}{MIP}\\ \cmidrule{3-8}\cmidrule{10-13}
Instance && Avg & Best & Time & AGap & BGap & $|M|$ && MIP & Time & MIPGap & $|M|$\\
\midrule
8\_5\_101 && 3491.7 & 3509.0 & 233.9 & 0.49 & 0.0 & 52.6 && 3509 & 942 & 0 & 44\\
8\_4\_160 && 4229.0 & 4229.0 & 256.4 & 0.87 & 0.87 & 60.0 && 4266 & 459 & 0 & 53\\
8\_2\_119 && 1460.0 & 1460.0 & 100.2 & 0.0 & 0.0 & 22.0 && 1460 & 0.5 & 0 & 22\\
8\_5\_72 && 2705.2 & 2734.0 & 172.6 & 1.05 & 0.0 & 35.2 && 2734 & 2 & 0 & 35\\
8\_2\_109 && 2348.0 & 2360.0 & 123.0 & 0.51 & 0.0 & 29.2 && 2360 & 14 & 0 & 30\\
8\_4\_172 && 5218.0 & 5250.0 & 288.9 & 0.61 & 0.0 & 60.6 && 5250 & 753 & 0 & 61\\
8\_5\_124 && 3565.0 & 3565.0 & 207.6 & 0.0 & 0.0 & 48.0 && 3565 & 82 & 0 & 40\\
8\_5\_145 && 4678.4 & 4680.0 & 249.2 & 0.88 & 0.85 & 67.4 && 4720 & 21600 & 3.97 & 69\\
8\_5\_80 && 2700.7 & 2707.0 & 176.2 & 0.64 & 0.4 & 33.2 && 2718 & 61 & 0 & 33\\
8\_6\_136 && 4315.0 & 4315.0 & 309.5 & 0.46 & 0.46 & 76.0 && 4335 & 21600 & 0.62 & 75\\
18\_10\_567 && 9411.8 & 9525.0 & 1077.5 & 1.81 & 0.63 & 132.2 && 9585 & 21600 & 6.17 & 135\\
18\_9\_548 && 10168.4 & 10280.0 & 952.7 & 0.56 & -0.53 & 142.1 && 10226 & 21600 & 8.51 & 141\\
18\_4\_412 && 3428.0 & 3440.0 & 331.1 & 0.35 & 0.0 & 38.4 && 3440 & 12560 & 0 & 39\\
18\_6\_440 && 6049.4 & 6137.0 & 453.6 & 1.43 & 0.0 & 70.1 && 6137 & 21600 & 4.31 & 70\\
18\_4\_404 && 4924.0 & 4980.0 & 391.3 & 1.03 & -0.1 & 65.1 && 4975 & 21600 & 4.32 & 66\\
18\_8\_514 && 9312.4 & 9352.0 & 644.7 & 0.74 & 0.32 & 106.5 && 9382 & 21600 & 4.12 & 108\\
18\_10\_710 && 9011.6 & 9130.0 & 1168.4 & 0.31 & -1.0 & 120.6 && 9040 & 21600 & 10.55 & 119\\
18\_8\_557 && 8587.4 & 8620.0 & 825.5 & 1.36 & 0.99 & 114.1 && 8706 & 21600 & 8.66 & 116\\
18\_4\_363 && 3337.0 & 3337.0 & 301.6 & 0.0 & 0.0 & 40.0 && 3337 & 43 & 0 & 40\\
18\_9\_368 && 6792.0 & 6849.0 & 590.2 & 0.03 & -0.81 & 81.8 && 6794 & 21600 & 7.85 & 81\\
\midrule
Average &&&&& 0.66 & 0.1 &&&&&\\
\bottomrule
\end{tabular}
\end{table}

\begin{table}[tbh]
\caption{Results for small and medium-sized instances (1,000 ALNS iterations, $\epsilon_3 = 0$)\label{secondstart}}
\scriptsize
\centering
\begin{tabular}{lp{3pt}rrrrrrp{3pt}rrrl}
\toprule
&&\multicolumn{6}{c}{LNS}&&\multicolumn{4}{c}{MIP}\\ \cmidrule{3-8}\cmidrule{10-13}
Instance && Avg & Best & Time & AGap & BGap & $|M|$ && MIP & Time & MIPGap & $|M|$\\
\midrule
8\_5\_101 && 3405.4 & 3429.0 & 7.4 & 2.95 & 2.28 & 51.7 && 3509 & 942 & 0 & 44\\
8\_4\_160 && 4165.0 & 4200.0 & 7.7 & 2.37 & 1.55 & 61.1 && 4266 & 459 & 0 & 53\\
8\_2\_119 && 1454.0 & 1460.0 & 4.7 & 0.41 & 0.0 & 22.0 && 1460 & 0.5 & 0 & 22\\
8\_5\_72 && 2673.7 & 2731.0 & 6.4 & 2.21 & 0.11 & 35.0 && 2734 & 2 & 0 & 35\\
8\_2\_109 && 2308.0 & 2350.0 & 5.1 & 2.2 & 0.42 & 28.9 && 2360 & 14 & 0 & 30\\
8\_4\_172 && 4990.0 & 5120.0 & 8.1 & 4.95 & 2.48 & 57.5 && 5250 & 753 & 0 & 61\\
8\_5\_124 && 3455.0 & 3565.0 & 6.8 & 3.09 & 0.0 & 47.6 && 3565 & 82 & 0 & 40\\
8\_5\_145 && 4605.0 & 4650.0 & 8.3 & 2.44 & 1.48 & 66.2 && 4720 & 21600 & 3.97 & 69\\
8\_5\_80 && 2630.4 & 2667.0 & 6.5 & 3.22 & 1.88 & 33.1 && 2718 & 61 & 0 & 33\\
8\_6\_136 && 4238.0 & 4295.0 & 8.9 & 2.24 & 0.92 & 76.0 && 4335 & 21600 & 0.62 & 75\\
18\_10\_567 && 8988.9 & 9085.0 & 18.2 & 6.22 & 5.22 & 127.2 && 9585 & 21600 & 6.17 & 135\\
18\_9\_548 && 9802.4 & 9912.0 & 16.1 & 4.14 & 3.07 & 137.8 && 10226 & 21600 & 8.51 & 141\\
18\_4\_412 && 3369.0 & 3420.0 & 8.0 & 2.06 & 0.58 & 37.9 && 3440 & 12560 & 0 & 39\\
18\_6\_440 && 5935.1 & 6007.0 & 9.8 & 3.29 & 2.12 & 69.1 && 6137 & 21600 & 4.31 & 70\\
18\_4\_404 && 4818.5 & 4870.0 & 9.0 & 3.15 & 2.11 & 63.5 && 4975 & 21600 & 4.32 & 66\\
18\_8\_514 && 8957.4 & 9150.0 & 12.8 & 4.53 & 2.47 & 102.0 && 9382 & 21600 & 4.12 & 108\\
18\_10\_710 && 8600.9 & 8758.0 & 18.7 & 4.86 & 3.12 & 116.7 && 9040 & 21600 & 10.55 & 119\\
18\_8\_557 && 8230.5 & 8330.0 & 14.4 & 5.46 & 4.32 & 110.7 && 8706 & 21600 & 8.66 & 116\\
18\_4\_363 && 3337.0 & 3337.0 & 7.3 & 0.0 & 0.0 & 40.0 && 3337 & 43 & 0 & 40\\
18\_9\_368 && 6622.7 & 6695.0 & 11.2 & 2.52 & 1.46 & 80.6 && 6794 & 21600 & 7.85 & 81\\
\midrule
Average &&&&& 3.12 & 1.78 &&&&&\\
\bottomrule
\end{tabular}
\end{table}
\begin{table}[tbh]
\caption{Results for small and medium-sized instances (10,000 ALNS iterations, $\epsilon_3 = 0$)}
\scriptsize
\centering
\begin{tabular}{lp{3pt}rrrrrrp{3pt}rrrl}
\toprule
&&\multicolumn{6}{c}{LNS}&&\multicolumn{4}{c}{MIP}\\ \cmidrule{3-8}\cmidrule{10-13}
Instance && Avg & Best & Time & AGap & BGap & $|M|$ && MIP & Time & MIPGap & $|M|$\\
\midrule
8\_5\_101 && 3457.4 & 3499.0 & 30.4 & 1.47 & 0.28 & 52.8 && 3509 & 942 & 0 & 44\\
8\_4\_160 && 4200.9 & 4229.0 & 35.0 & 1.53 & 0.87 & 61.1 && 4266 & 459 & 0 & 53\\
8\_2\_119 && 1460.0 & 1460.0 & 16.3 & 0.0 & 0.0 & 22.0 && 1460 & 0.5 & 0 & 22\\
8\_5\_72 && 2691.4 & 2731.0 & 24.2 & 1.56 & 0.11 & 35.0 && 2734 & 2 & 0 & 35\\
8\_2\_109 && 2337.0 & 2360.0 & 19.7 & 0.97 & 0.0 & 29.2 && 2360 & 14 & 0 & 30\\
8\_4\_172 && 5111.0 & 5250.0 & 37.5 & 2.65 & 0.0 & 59.1 && 5250 & 753 & 0 & 61\\
8\_5\_124 && 3531.0 & 3565.0 & 26.4 & 0.95 & 0.0 & 48.2 && 3565 & 82 & 0 & 40\\
8\_5\_145 && 4642.8 & 4680.0 & 34.9 & 1.64 & 0.85 & 67.2 && 4720 & 21600 & 3.97 & 69\\
8\_5\_80 && 2679.9 & 2707.0 & 24.4 & 1.4 & 0.4 & 33.4 && 2718 & 61 & 0 & 33\\
8\_6\_136 && 4295.0 & 4315.0 & 40.2 & 0.92 & 0.46 & 75.3 && 4335 & 21600 & 0.62 & 75\\
18\_10\_567 && 9247.3 & 9395.0 & 120.7 & 3.52 & 1.98 & 130.0 && 9585 & 21600 & 6.17 & 135\\
18\_9\_548 && 10034.9 & 10102.0 & 103.5 & 1.87 & 1.21 & 139.9 && 10226 & 21600 & 8.51 & 141\\
18\_4\_412 && 3420.0 & 3440.0 & 40.5 & 0.58 & 0.0 & 38.4 && 3440 & 12560 & 0 & 39\\
18\_6\_440 && 6031.0 & 6107.0 & 52.2 & 1.73 & 0.49 & 70.1 && 6137 & 21600 & 4.31 & 70\\
18\_4\_404 && 4894.0 & 4975.0 & 46.9 & 1.63 & 0.0 & 64.3 && 4975 & 21600 & 4.32 & 66\\
18\_8\_514 && 9209.4 & 9338.0 & 76.8 & 1.84 & 0.47 & 106.2 && 9382 & 21600 & 4.12 & 108\\
18\_10\_710 && 8837.9 & 9005.0 & 128.0 & 2.24 & 0.39 & 119.5 && 9040 & 21600 & 10.55 & 119\\
18\_8\_557 && 8499.4 & 8630.0 & 91.6 & 2.37 & 0.87 & 112.9 && 8706 & 21600 & 8.66 & 116\\
18\_4\_363 && 3337.0 & 3337.0 & 38.2 & 0.0 & 0.0 & 40.0 && 3337 & 43 & 0 & 40\\
18\_9\_368 && 6713.7 & 6759.0 & 66.5 & 1.18 & 0.52 & 80.6 && 6794 & 21600 & 7.85 & 81\\
\midrule
Average &&&&& 1.5 & 0.44 &&&&&\\
\bottomrule
\end{tabular}
\end{table}
\begin{table}[tbh]
\caption{Results for small and medium-sized instances (25,000 ALNS iterations, $\epsilon_3 = 0$)}
\scriptsize
\centering
\begin{tabular}{lp{3pt}rrrrrrp{3pt}rrrl}
\toprule
&&\multicolumn{6}{c}{LNS}&&\multicolumn{4}{c}{MIP}\\ \cmidrule{3-8}\cmidrule{10-13}
Instance && Avg & Best & Time & AGap & BGap & $|M|$ && MIP & Time & MIPGap & $|M|$\\
\midrule
8\_5\_101 && 3478.7 & 3499.0 & 65.1 & 0.86 & 0.28 & 52.7 && 3509 & 942 & 0 & 44\\
8\_4\_160 && 4202.9 & 4229.0 & 74.8 & 1.48 & 0.87 & 60.8 && 4266 & 459 & 0 & 53\\
8\_2\_119 && 1460.0 & 1460.0 & 32.2 & 0.0 & 0.0 & 22.0 && 1460 & 0.5 & 0 & 22\\
8\_5\_72 && 2691.4 & 2731.0 & 48.4 & 1.56 & 0.11 & 35.0 && 2734 & 2 & 0 & 35\\
8\_2\_109 && 2341.0 & 2360.0 & 39.0 & 0.81 & 0.0 & 29.2 && 2360 & 14 & 0 & 30\\
8\_4\_172 && 5175.0 & 5250.0 & 80.9 & 1.43 & 0.0 & 60.0 && 5250 & 753 & 0 & 61\\
8\_5\_124 && 3545.0 & 3565.0 & 55.5 & 0.56 & 0.0 & 48.3 && 3565 & 82 & 0 & 40\\
8\_5\_145 && 4643.8 & 4680.0 & 72.1 & 1.61 & 0.85 & 67.1 && 4720 & 21600 & 3.97 & 69\\
8\_5\_80 && 2688.5 & 2707.0 & 49.1 & 1.09 & 0.4 & 33.5 && 2718 & 61 & 0 & 33\\
8\_6\_136 && 4301.5 & 4315.0 & 88.7 & 0.77 & 0.46 & 75.6 && 4335 & 21600 & 0.62 & 75\\
18\_10\_567 && 9301.5 & 9405.0 & 284.5 & 2.96 & 1.88 & 131.3 && 9585 & 21600 & 6.17 & 135\\
18\_9\_548 && 10074.0 & 10200.0 & 244.8 & 1.49 & 0.25 & 140.5 && 10226 & 21600 & 8.51 & 141\\
18\_4\_412 && 3426.0 & 3440.0 & 88.3 & 0.41 & 0.0 & 38.6 && 3440 & 12560 & 0 & 39\\
18\_6\_440 && 6031.0 & 6107.0 & 117.7 & 1.73 & 0.49 & 70.1 && 6137 & 21600 & 4.31 & 70\\
18\_4\_404 && 4899.5 & 4975.0 & 103.7 & 1.52 & 0.0 & 64.5 && 4975 & 21600 & 4.32 & 66\\
18\_8\_514 && 9264.6 & 9352.0 & 175.8 & 1.25 & 0.32 & 106.7 && 9382 & 21600 & 4.12 & 108\\
18\_10\_710 && 8881.3 & 9035.0 & 320.4 & 1.76 & 0.06 & 119.3 && 9040 & 21600 & 10.55 & 119\\
18\_8\_557 && 8525.6 & 8630.0 & 214.7 & 2.07 & 0.87 & 112.4 && 8706 & 21600 & 8.66 & 116\\
18\_4\_363 && 3337.0 & 3337.0 & 87.3 & 0.0 & 0.0 & 40.0 && 3337 & 43 & 0 & 40\\
18\_9\_368 && 6731.8 & 6759.0 & 153.2 & 0.92 & 0.52 & 81.1 && 6794 & 21600 & 7.85 & 81\\
\midrule
Average &&&&& 1.21 & 0.37 &&&&&\\
\bottomrule
\end{tabular}
\end{table}
\begin{table}[tbh]
\caption{Results for small and medium-sized instances (100,000 ALNS iterations, $\epsilon_3 = 0$)\label{tableend}}
\scriptsize
\centering
\begin{tabular}{lp{3pt}rrrrrrp{3pt}rrrl}
\toprule
&&\multicolumn{6}{c}{LNS}&&\multicolumn{4}{c}{MIP}\\ \cmidrule{3-8}\cmidrule{10-13}
Instance && Avg & Best & Time & AGap & BGap & $|M|$ && MIP & Time & MIPGap & $|M|$\\
\midrule
8\_5\_101 && 3492.3 & 3509.0 & 226.7 & 0.48 & 0.0 & 52.2 && 3509 & 942 & 0 & 44\\
8\_4\_160 && 4202.9 & 4229.0 & 271.9 & 1.48 & 0.87 & 60.8 && 4266 & 459 & 0 & 53\\
8\_2\_119 && 1460.0 & 1460.0 & 108.6 & 0.0 & 0.0 & 22.0 && 1460 & 0.5 & 0 & 22\\
8\_5\_72 && 2691.4 & 2731.0 & 166.7 & 1.56 & 0.11 & 35.0 && 2734 & 2 & 0 & 35\\
8\_2\_109 && 2350.0 & 2360.0 & 132.8 & 0.42 & 0.0 & 29.2 && 2360 & 14 & 0 & 30\\
8\_4\_172 && 5211.0 & 5250.0 & 285.2 & 0.74 & 0.0 & 60.5 && 5250 & 753 & 0 & 61\\
8\_5\_124 && 3551.0 & 3565.0 & 190.3 & 0.39 & 0.0 & 48.4 && 3565 & 82 & 0 & 40\\
8\_5\_145 && 4643.8 & 4680.0 & 253.3 & 1.61 & 0.85 & 67.1 && 4720 & 21600 & 3.97 & 69\\
8\_5\_80 && 2697.0 & 2718.0 & 166.1 & 0.77 & 0.0 & 33.4 && 2718 & 61 & 0 & 33\\
8\_6\_136 && 4308.0 & 4320.0 & 328.8 & 0.62 & 0.35 & 75.7 && 4335 & 21600 & 0.62 & 75\\
18\_10\_567 && 9367.0 & 9575.0 & 1117.0 & 2.27 & 0.1 & 131.8 && 9585 & 21600 & 6.17 & 135\\
18\_9\_548 && 10123.2 & 10200.0 & 935.0 & 1.01 & 0.25 & 139.0 && 10226 & 21600 & 8.51 & 141\\
18\_4\_412 && 3426.0 & 3440.0 & 326.3 & 0.41 & 0.0 & 38.6 && 3440 & 12560 & 0 & 39\\
18\_6\_440 && 6041.0 & 6127.0 & 428.4 & 1.56 & 0.16 & 70.0 && 6137 & 21600 & 4.31 & 70\\
18\_4\_404 && 4918.0 & 4980.0 & 389.3 & 1.15 & -0.1 & 65.0 && 4975 & 21600 & 4.32 & 66\\
18\_8\_514 && 9297.4 & 9362.0 & 633.5 & 0.9 & 0.21 & 107.0 && 9382 & 21600 & 4.12 & 108\\
18\_10\_710 && 8947.3 & 9065.0 & 1246.6 & 1.03 & -0.28 & 120.4 && 9040 & 21600 & 10.55 & 119\\
18\_8\_557 && 8549.6 & 8630.0 & 803.4 & 1.8 & 0.87 & 112.7 && 8706 & 21600 & 8.66 & 116\\
18\_4\_363 && 3337.0 & 3337.0 & 322.1 & 0.0 & 0.0 & 40.0 && 3337 & 43 & 0 & 40\\
18\_9\_368 && 6735.6 & 6767.0 & 568.9 & 0.86 & 0.4 & 81.0 && 6794 & 21600 & 7.85 & 81\\
\midrule
Average &&&&& 0.95 & 0.19 &&&&&\\
\bottomrule
\end{tabular}
\end{table}

\begin{table}[tbh]
\caption{ALNS results for setting \tt{Priority}, using weight set \{1,0,0,0,0\} and 100,000 iterations. \label{tbl:priority}}
\scriptsize
\centering
\begin{tabular}{lp{3pt}rrrrrr}
\toprule
Instance && Priority & Penalty & Groups & Consec & Workload & OF\\
\midrule
8\_5\_101 && 3496.0 & -15.8 & 4.18 & 0.0 & 0.9 & 3480.2\\
8\_4\_160 && 4257.0 & -20.9 & 3.23 & 0.0 & 0.91 & 4236.1\\
8\_2\_119 && 1460.0 & -3.0 & 2.0 & 0.0 & 0.88 & 1457.0\\
8\_5\_72 && 2816.0 & -155.4 & 3.32 & 0.0 & 0.88 & 2660.6\\
8\_2\_109 && 2341.0 & 0.0 & 3.65 & 0.0 & 0.87 & 2341.0\\
8\_4\_172 && 5219.0 & 0.0 & 2.5 & 0.0 & 0.87 & 5219.0\\
8\_5\_124 && 3548.0 & -11.8 & 2.4 & 0.0 & 0.85 & 3536.2\\
8\_5\_145 && 4702.0 & -27.7 & 1.66 & 0.0 & 0.89 & 4674.3\\
8\_5\_80 && 2814.0 & -151.5 & 2.82 & 0.8 & 0.89 & 2662.5\\
8\_6\_136 && 4321.0 & -7.5 & 2.93 & 0.0 & 0.92 & 4313.5\\
18\_10\_567 && 9426.0 & -14.1 & 2.66 & 2.6 & 0.9 & 9411.9\\
18\_9\_548 && 10153.0 & -9.8 & 2.76 & 2.7 & 0.9 & 10143.2\\
18\_4\_412 && 3434.0 & 0.0 & 1.35 & 3.9 & 0.92 & 3434.0\\
18\_6\_440 && 6049.0 & -33.0 & 2.48 & 6.3 & 0.88 & 6016.0\\
18\_4\_404 && 4958.0 & -75.0 & 3.25 & 0.1 & 0.9 & 4883.0\\
18\_8\_514 && 9684.0 & -517.3 & 1.74 & 3.1 & 0.86 & 9166.7\\
18\_10\_710 && 9090.0 & -117.3 & 2.13 & 1.7 & 0.87 & 8972.7\\
18\_8\_557 && 8610.0 & -3.8 & 2.38 & 0.8 & 0.9 & 8606.2\\
18\_4\_363 && 3420.0 & -83.0 & 1.75 & 1.2 & 0.87 & 3337.0\\
18\_9\_368 && 6836.0 & -93.5 & 2.16 & 4.5 & 0.89 & 6742.5\\
24\_45\_1072 && 31824.0 & -495.8 & 2.71 & 8.2 & 0.85 & 31328.2\\
24\_38\_906 && 32734.0 & -186.5 & 2.29 & 10.6 & 0.87 & 32547.5\\
24\_45\_621 && 32865.0 & -254.5 & 1.63 & 15.8 & 0.83 & 32610.5\\
24\_43\_767 && 35408.0 & -476.3 & 1.75 & 18.3 & 0.85 & 34931.7\\
24\_42\_749 && 36360.0 & -313.6 & 2.1 & 9.9 & 0.84 & 36046.4\\
24\_40\_825 && 38155.0 & -296.5 & 2.32 & 6.4 & 0.84 & 37858.5\\
24\_45\_1441 && 39270.0 & -150.9 & 1.9 & 4.7 & 0.84 & 39119.1\\
24\_45\_1076 && 36358.0 & -182.9 & 2.25 & 5.2 & 0.87 & 36175.1\\
24\_44\_610 && 32844.0 & -241.0 & 2.14 & 9.3 & 0.84 & 32603.0\\
24\_52\_662 && 36459.0 & -690.6 & 1.7 & 15.0 & 0.84 & 35768.4\\
72\_230\_2505 && 123114.0 & -1261.3 & 2.33 & 28.7 & 0.82 & 121852.7\\
72\_211\_2438 && 130057.0 & -676.6 & 2.09 & 37.2 & 0.83 & 129380.4\\
72\_224\_2436 && 142137.0 & -991.9 & 1.97 & 40.2 & 0.81 & 141145.1\\
72\_211\_2400 && 145159.0 & -3375.0 & 1.98 & 45.6 & 0.82 & 141784.0\\
72\_216\_2027 && 128556.0 & -838.7 & 2.19 & 24.0 & 0.8 & 127717.3\\
72\_230\_3127 && 155038.0 & -2106.8 & 2.36 & 21.3 & 0.81 & 152931.2\\
72\_213\_3244 && 145748.0 & -1056.9 & 2.23 & 20.7 & 0.83 & 144691.1\\
72\_218\_2456 && 126024.0 & -1151.5 & 2.04 & 31.1 & 0.82 & 124872.5\\
72\_234\_2543 && 146958.0 & -1800.5 & 2.26 & 34.0 & 0.82 & 145157.5\\
72\_180\_2258 && 93309.0 & -1428.8 & 1.98 & 25.9 & 0.81 & 91880.2\\
\midrule
Average && 44875.28 & -482.92 & 2.34 & 11.0 & 0.86 & 44392.35\\
\bottomrule
\end{tabular}
\end{table}
\begin{table}[tbh]
\caption{ALNS results for setting \tt{Penalty}, using weight set \{1,100,0,0,0\} and 100,000 iterations \label{tbl:penalty}}
\scriptsize
\centering
\begin{tabular}{lp{3pt}rrrrrr}
\toprule
Instance && Priority & Penalty & Groups & Consec & Workload & OF\\
\midrule
8\_5\_101 && 3464.0 & 0.0 & 4.42 & 0.0 & 0.89 & 3464.0\\
8\_4\_160 && 4200.0 & 0.0 & 3.0 & 0.0 & 0.92 & 4200.0\\
8\_2\_119 && 1460.0 & 0.0 & 2.0 & 0.0 & 0.88 & 1460.0\\
8\_5\_72 && 2650.0 & 0.0 & 2.88 & 0.0 & 0.86 & 2650.0\\
8\_2\_109 && 2356.0 & 0.0 & 3.75 & 0.0 & 0.86 & 2356.0\\
8\_4\_172 && 5240.0 & 0.0 & 2.5 & 0.0 & 0.87 & 5240.0\\
8\_5\_124 && 3480.0 & 0.0 & 2.46 & 0.0 & 0.86 & 3480.0\\
8\_5\_145 && 4680.0 & 0.0 & 1.6 & 0.0 & 0.89 & 4680.0\\
8\_5\_80 && 2510.0 & 0.0 & 2.74 & 0.4 & 0.89 & 2510.0\\
8\_6\_136 && 4288.0 & 0.0 & 3.0 & 0.0 & 0.92 & 4288.0\\
18\_10\_567 && 9311.0 & 0.0 & 2.55 & 1.6 & 0.89 & 9311.0\\
18\_9\_548 && 10162.0 & 0.0 & 2.74 & 2.3 & 0.9 & 10162.0\\
18\_4\_412 && 3434.0 & 0.0 & 1.35 & 2.9 & 0.92 & 3434.0\\
18\_6\_440 && 6018.0 & 0.0 & 2.43 & 7.5 & 0.88 & 6018.0\\
18\_4\_404 && 4886.0 & 0.0 & 2.55 & 0.2 & 0.91 & 4886.0\\
18\_8\_514 && 9256.0 & 0.0 & 1.88 & 3.3 & 0.87 & 9256.0\\
18\_10\_710 && 8959.0 & 0.0 & 2.22 & 1.7 & 0.88 & 8959.0\\
18\_8\_557 && 8630.0 & 0.0 & 2.44 & 0.7 & 0.9 & 8630.0\\
18\_4\_363 && 3040.0 & 0.0 & 1.5 & 1.4 & 0.91 & 3040.0\\
18\_9\_368 && 6674.0 & 0.0 & 2.42 & 6.5 & 0.9 & 6674.0\\
24\_45\_1072 && 31328.0 & 0.0 & 2.72 & 5.4 & 0.86 & 31328.0\\
24\_38\_906 && 32452.0 & 0.0 & 2.63 & 9.4 & 0.88 & 32452.0\\
24\_45\_621 && 32668.0 & 0.0 & 1.67 & 14.7 & 0.84 & 32668.0\\
24\_43\_767 && 35086.0 & 0.0 & 1.69 & 17.1 & 0.84 & 35086.0\\
24\_42\_749 && 35799.0 & 0.0 & 2.09 & 9.1 & 0.84 & 35799.0\\
24\_40\_825 && 37765.0 & 0.0 & 2.31 & 5.8 & 0.85 & 37765.0\\
24\_45\_1441 && 38991.0 & 0.0 & 1.89 & 5.7 & 0.84 & 38991.0\\
24\_45\_1076 && 36187.0 & 0.0 & 2.23 & 5.8 & 0.87 & 36187.0\\
24\_44\_610 && 31753.0 & 0.0 & 1.85 & 10.2 & 0.83 & 31753.0\\
24\_52\_662 && 35600.0 & 0.0 & 1.81 & 17.2 & 0.84 & 35600.0\\
72\_230\_2505 && 121500.0 & 0.0 & 2.37 & 29.3 & 0.83 & 121500.0\\
72\_211\_2438 && 128646.0 & 0.0 & 2.1 & 34.4 & 0.83 & 128646.0\\
72\_224\_2436 && 141154.0 & 0.0 & 1.93 & 46.2 & 0.81 & 141154.0\\
72\_211\_2400 && 140377.0 & 0.0 & 2.02 & 38.3 & 0.82 & 140377.0\\
72\_216\_2027 && 126875.0 & 0.0 & 2.22 & 25.0 & 0.8 & 126875.0\\
72\_230\_3127 && 152438.0 & 0.0 & 2.45 & 20.0 & 0.82 & 152438.0\\
72\_213\_3244 && 141764.0 & 0.0 & 2.28 & 18.6 & 0.83 & 141764.0\\
72\_218\_2456 && 124510.0 & -0.1 & 2.06 & 27.5 & 0.83 & 124509.9\\
72\_234\_2543 && 140511.0 & 0.0 & 2.06 & 29.8 & 0.81 & 140511.0\\
72\_180\_2258 && 89846.0 & 0.0 & 1.95 & 29.9 & 0.82 & 89846.0\\
\midrule
Average && 43998.7 & -0.0 & 2.32 & 10.7 & 0.86 & 43998.7\\
\bottomrule
\end{tabular}
\end{table}
\begin{table}[tbh]
\caption{ALNS results for setting \tt{TaskGroups}, using weight set \{1,1,-10,0,0\} and 100,000 iterations. \label{tbl:parallel}}
\scriptsize
\centering
\begin{tabular}{lp{3pt}rrrrrr}
\toprule
Instance && Priority & Penalty & Groups & Consec & Workload & OF\\
\midrule
8\_5\_101 && 3492.0 & -3.8 & 4.0 & 0.0 & 0.89 & 3488.2\\
8\_4\_160 && 4240.0 & -11.0 & 3.25 & 0.0 & 0.91 & 4229.0\\
8\_2\_119 && 1460.0 & 0.0 & 2.0 & 0.0 & 0.88 & 1460.0\\
8\_5\_72 && 2734.0 & -24.7 & 2.56 & 0.0 & 0.84 & 2709.3\\
8\_2\_109 && 2345.0 & 0.0 & 3.3 & 0.0 & 0.87 & 2345.0\\
8\_4\_172 && 5234.0 & 0.0 & 2.15 & 0.0 & 0.87 & 5234.0\\
8\_5\_124 && 3570.0 & -5.0 & 2.4 & 0.0 & 0.86 & 3565.0\\
8\_5\_145 && 4679.0 & -1.9 & 1.62 & 0.0 & 0.89 & 4677.1\\
8\_5\_80 && 2769.0 & -65.2 & 2.58 & 0.6 & 0.87 & 2703.8\\
8\_6\_136 && 4318.0 & -4.5 & 2.75 & 0.0 & 0.92 & 4313.5\\
18\_10\_567 && 9430.0 & -11.7 & 2.6 & 1.7 & 0.89 & 9418.3\\
18\_9\_548 && 10157.0 & -0.8 & 2.77 & 2.3 & 0.9 & 10156.2\\
18\_4\_412 && 3432.0 & 0.0 & 1.3 & 4.2 & 0.92 & 3432.0\\
18\_6\_440 && 6060.0 & -11.7 & 2.5 & 7.7 & 0.87 & 6048.3\\
18\_4\_404 && 4953.0 & -29.0 & 2.98 & 0.0 & 0.91 & 4924.0\\
18\_8\_514 && 9440.0 & -99.4 & 1.91 & 3.3 & 0.86 & 9340.6\\
18\_10\_710 && 9082.0 & -61.3 & 2.18 & 1.7 & 0.87 & 9020.7\\
18\_8\_557 && 8578.0 & -1.6 & 2.36 & 0.5 & 0.9 & 8576.4\\
18\_4\_363 && 3420.0 & -83.0 & 1.75 & 1.2 & 0.86 & 3337.0\\
18\_9\_368 && 6825.0 & -35.8 & 2.16 & 6.5 & 0.89 & 6789.2\\
24\_45\_1072 && 31630.0 & -141.0 & 2.66 & 5.9 & 0.86 & 31489.0\\
24\_38\_906 && 32608.0 & -29.0 & 2.5 & 8.2 & 0.88 & 32579.0\\
24\_45\_621 && 32788.0 & -76.1 & 1.66 & 14.6 & 0.83 & 32711.9\\
24\_43\_767 && 35352.0 & -73.7 & 1.73 & 19.9 & 0.84 & 35278.3\\
24\_42\_749 && 36239.0 & -170.4 & 2.13 & 9.6 & 0.84 & 36068.6\\
24\_40\_825 && 38075.0 & -117.7 & 2.27 & 7.6 & 0.83 & 37957.3\\
24\_45\_1441 && 39238.0 & -97.3 & 1.88 & 5.7 & 0.84 & 39140.7\\
24\_45\_1076 && 36301.0 & -58.8 & 2.26 & 5.6 & 0.86 & 36242.2\\
24\_44\_610 && 32804.0 & -163.2 & 2.02 & 9.1 & 0.83 & 32640.8\\
24\_52\_662 && 36167.0 & -216.0 & 1.76 & 15.3 & 0.84 & 35951.0\\
72\_230\_2505 && 122365.0 & -320.5 & 2.32 & 28.9 & 0.83 & 122044.5\\
72\_211\_2438 && 129745.0 & -316.4 & 2.09 & 39.0 & 0.83 & 129428.6\\
72\_224\_2436 && 141899.0 & -374.8 & 1.93 & 45.8 & 0.81 & 141524.2\\
72\_211\_2400 && 144527.0 & -1641.1 & 1.98 & 41.6 & 0.82 & 142885.9\\
72\_216\_2027 && 128356.0 & -388.4 & 2.2 & 27.2 & 0.8 & 127967.6\\
72\_230\_3127 && 154881.0 & -913.2 & 2.35 & 23.4 & 0.81 & 153967.8\\
72\_213\_3244 && 145543.0 & -695.8 & 2.19 & 20.7 & 0.82 & 144847.2\\
72\_218\_2456 && 125263.0 & -277.8 & 2.02 & 32.6 & 0.83 & 124985.2\\
72\_234\_2543 && 146888.0 & -1120.7 & 2.22 & 27.6 & 0.82 & 145767.3\\
72\_180\_2258 && 92247.0 & -600.2 & 1.92 & 29.0 & 0.81 & 91646.8\\
\midrule
Average && 44728.35 & -206.06 & 2.28 & 11.18 & 0.86 & 44522.29\\
\bottomrule
\end{tabular}
\end{table}
\begin{table}[tbh]
\caption{ALNS results for setting \tt{Workload}, using weight set \{1,1,0,0,10000\} and 100,000 iterations. \label{tbl:percent}}
\scriptsize
\centering
\begin{tabular}{lp{3pt}rrrrrr}
\toprule
Instance && Priority & Penalty & Groups & Consec & Workload & OF\\
\midrule
8\_5\_101 && 3435.0 & -6.4 & 4.36 & 0.0 & 0.94 & 3428.6\\
8\_4\_160 && 4163.0 & -18.7 & 3.25 & 0.0 & 0.93 & 4144.3\\
8\_2\_119 && 1355.0 & -2.0 & 2.0 & 0.0 & 0.96 & 1353.0\\
8\_5\_72 && 2656.0 & -22.5 & 3.18 & 0.0 & 0.93 & 2633.5\\
8\_2\_109 && 2272.0 & 0.0 & 3.4 & 0.0 & 0.96 & 2272.0\\
8\_4\_172 && 5061.0 & 0.0 & 2.62 & 0.0 & 0.92 & 5061.0\\
8\_5\_124 && 3420.0 & -5.0 & 2.56 & 0.0 & 0.89 & 3415.0\\
8\_5\_145 && 4586.0 & 0.0 & 1.4 & 0.0 & 0.92 & 4586.0\\
8\_5\_80 && 2704.0 & -79.6 & 2.76 & 0.7 & 0.92 & 2624.4\\
8\_6\_136 && 4183.0 & 0.0 & 3.28 & 0.0 & 0.94 & 4183.0\\
18\_10\_567 && 9224.0 & -13.5 & 2.75 & 2.6 & 0.91 & 9210.5\\
18\_9\_548 && 9963.0 & -1.8 & 3.12 & 2.4 & 0.92 & 9961.2\\
18\_4\_412 && 3434.0 & 0.0 & 1.35 & 4.3 & 0.93 & 3434.0\\
18\_6\_440 && 5921.0 & -10.4 & 2.73 & 4.5 & 0.92 & 5910.6\\
18\_4\_404 && 4939.0 & -18.5 & 3.02 & 0.4 & 0.92 & 4920.5\\
18\_8\_514 && 9069.0 & -29.6 & 2.11 & 2.6 & 0.91 & 9039.4\\
18\_10\_710 && 8850.0 & -65.5 & 2.15 & 2.0 & 0.89 & 8784.5\\
18\_8\_557 && 8475.0 & -0.4 & 2.36 & 0.6 & 0.91 & 8474.6\\
18\_4\_363 && 3340.0 & -68.0 & 1.75 & 1.0 & 0.93 & 3272.0\\
18\_9\_368 && 6697.0 & -38.2 & 2.33 & 6.1 & 0.91 & 6658.8\\
24\_45\_1072 && 31413.0 & -139.5 & 2.72 & 8.3 & 0.86 & 31273.5\\
24\_38\_906 && 32427.0 & -38.1 & 2.52 & 8.6 & 0.88 & 32388.9\\
24\_45\_621 && 32659.0 & -76.9 & 1.69 & 13.2 & 0.84 & 32582.1\\
24\_43\_767 && 35094.0 & -51.3 & 1.73 & 20.4 & 0.85 & 35042.7\\
24\_42\_749 && 36053.0 & -239.9 & 2.16 & 8.9 & 0.85 & 35813.1\\
24\_40\_825 && 37760.0 & -133.7 & 2.39 & 4.8 & 0.85 & 37626.3\\
24\_45\_1441 && 39095.0 & -98.4 & 1.94 & 6.1 & 0.85 & 38996.6\\
24\_45\_1076 && 36057.0 & -60.2 & 2.2 & 5.7 & 0.87 & 35996.8\\
24\_44\_610 && 32615.0 & -162.2 & 2.03 & 7.9 & 0.84 & 32452.8\\
24\_52\_662 && 35937.0 & -165.9 & 1.84 & 16.3 & 0.85 & 35771.1\\
72\_230\_2505 && 122360.0 & -363.2 & 2.36 & 25.8 & 0.83 & 121996.8\\
72\_211\_2438 && 129625.0 & -253.7 & 2.08 & 37.1 & 0.83 & 129371.3\\
72\_224\_2436 && 141830.0 & -368.2 & 1.95 & 43.6 & 0.82 & 141461.8\\
72\_211\_2400 && 144502.0 & -1543.8 & 1.99 & 44.3 & 0.82 & 142958.2\\
72\_216\_2027 && 128236.0 & -432.7 & 2.16 & 23.5 & 0.81 & 127803.3\\
72\_230\_3127 && 154722.0 & -905.5 & 2.45 & 20.3 & 0.82 & 153816.5\\
72\_213\_3244 && 145427.0 & -640.8 & 2.19 & 22.5 & 0.83 & 144786.2\\
72\_218\_2456 && 125136.0 & -285.8 & 2.05 & 33.5 & 0.83 & 124850.2\\
72\_234\_2543 && 146546.0 & -1056.9 & 2.23 & 28.8 & 0.82 & 145489.1\\
72\_180\_2258 && 91809.0 & -404.8 & 1.95 & 28.8 & 0.82 & 91404.2\\
\midrule
Average && 44576.25 & -195.04 & 2.38 & 10.89 & 0.88 & 44381.21\\
\bottomrule
\end{tabular}
\end{table}
\begin{table}[tbh]
\caption{ALNS results for setting \tt{Control}, using weight set \{1,1,0,0,0\} and 100,000 iterations. \label{tbl:control}}
\scriptsize
\centering
\begin{tabular}{lp{3pt}rrrrrr}
\toprule
Instance && Priority & Penalty & Groups & Consec & Workload & OF\\
\midrule
8\_5\_101 && 3493.0 & -3.0 & 4.22 & 0.0 & 0.9 & 3490.0\\
8\_4\_160 && 4240.0 & -11.0 & 3.27 & 0.0 & 0.91 & 4229.0\\
8\_2\_119 && 1460.0 & 0.0 & 2.0 & 0.0 & 0.88 & 1460.0\\
8\_5\_72 && 2747.0 & -28.6 & 2.98 & 0.0 & 0.86 & 2718.4\\
8\_2\_109 && 2357.0 & 0.0 & 3.75 & 0.0 & 0.86 & 2357.0\\
8\_4\_172 && 5188.0 & 0.0 & 2.45 & 0.0 & 0.87 & 5188.0\\
8\_5\_124 && 3559.0 & -5.0 & 2.4 & 0.0 & 0.85 & 3554.0\\
8\_5\_145 && 4682.0 & -2.6 & 1.64 & 0.0 & 0.89 & 4679.4\\
8\_5\_80 && 2755.0 & -63.0 & 2.76 & 0.7 & 0.87 & 2692.0\\
8\_6\_136 && 4319.0 & -5.0 & 3.03 & 0.0 & 0.92 & 4314.0\\
18\_10\_567 && 9408.0 & -11.2 & 2.64 & 2.8 & 0.89 & 9396.8\\
18\_9\_548 && 10147.0 & -3.8 & 2.82 & 2.2 & 0.9 & 10143.2\\
18\_4\_412 && 3434.0 & 0.0 & 1.35 & 3.6 & 0.92 & 3434.0\\
18\_6\_440 && 6036.0 & -9.1 & 2.6 & 7.1 & 0.88 & 6026.9\\
18\_4\_404 && 4945.0 & -33.0 & 3.1 & 0.0 & 0.91 & 4912.0\\
18\_8\_514 && 9439.0 & -96.6 & 1.89 & 2.8 & 0.86 & 9342.4\\
18\_10\_710 && 9088.0 & -55.3 & 2.25 & 0.6 & 0.88 & 9032.7\\
18\_8\_557 && 8622.0 & -1.0 & 2.49 & 0.9 & 0.9 & 8621.0\\
18\_4\_363 && 3420.0 & -83.0 & 1.77 & 1.2 & 0.87 & 3337.0\\
18\_9\_368 && 6829.0 & -34.9 & 2.16 & 5.6 & 0.89 & 6794.1\\
24\_45\_1072 && 31669.0 & -149.7 & 2.61 & 7.4 & 0.86 & 31519.3\\
24\_38\_906 && 32617.0 & -34.9 & 2.43 & 8.8 & 0.87 & 32582.1\\
24\_45\_621 && 32847.0 & -93.0 & 1.67 & 17.2 & 0.83 & 32754.0\\
24\_43\_767 && 35291.0 & -85.8 & 1.71 & 19.4 & 0.84 & 35205.2\\
24\_42\_749 && 36242.0 & -182.5 & 2.1 & 8.6 & 0.84 & 36059.5\\
24\_40\_825 && 38064.0 & -188.2 & 2.29 & 6.5 & 0.83 & 37875.8\\
24\_45\_1441 && 39229.0 & -98.6 & 1.91 & 4.0 & 0.84 & 39130.4\\
24\_45\_1076 && 36382.0 & -63.8 & 2.23 & 6.1 & 0.86 & 36318.2\\
24\_44\_610 && 32774.0 & -154.1 & 2.05 & 10.1 & 0.83 & 32619.9\\
24\_52\_662 && 36156.0 & -177.5 & 1.76 & 16.4 & 0.84 & 35978.5\\
72\_230\_2505 && 122565.0 & -420.0 & 2.35 & 27.4 & 0.83 & 122145.0\\
72\_211\_2438 && 129817.0 & -317.1 & 2.14 & 37.0 & 0.83 & 129499.9\\
72\_224\_2436 && 141791.0 & -378.8 & 1.95 & 41.6 & 0.81 & 141412.2\\
72\_211\_2400 && 144650.0 & -1632.4 & 1.98 & 45.0 & 0.82 & 143017.6\\
72\_216\_2027 && 128449.0 & -449.7 & 2.16 & 27.5 & 0.8 & 127999.3\\
72\_230\_3127 && 154951.0 & -946.6 & 2.44 & 19.6 & 0.82 & 154004.4\\
72\_213\_3244 && 145484.0 & -634.4 & 2.19 & 20.1 & 0.83 & 144849.6\\
72\_218\_2456 && 125358.0 & -276.8 & 2.05 & 28.4 & 0.83 & 125081.2\\
72\_234\_2543 && 146766.0 & -1096.9 & 2.21 & 32.2 & 0.82 & 145669.1\\
72\_180\_2258 && 92074.0 & -435.0 & 1.96 & 31.4 & 0.81 & 91639.0\\
\midrule
Average && 44733.6 & -206.55 & 2.34 & 11.05 & 0.86 & 44527.05\\
\bottomrule
\end{tabular}
\end{table}
\begin{table}[tbh]
\caption{ALNS+ results for setting \tt{AllObjectives}, using weight set \{1,1,-1,1,1000\} and 100,000 iterations. \label{tbl:allobj}}
\scriptsize
\centering
\begin{tabular}{lp{3pt}rrrrrr}
\toprule
Instance && Priority & Penalty & Groups & Consec & Workload & OF\\
\midrule
8\_5\_101 && 3493.0 & -6.3 & 4.26 & 0.0 & 0.91 & 3486.7\\
8\_4\_160 && 4239.0 & -11.0 & 3.25 & 0.0 & 0.92 & 4228.0\\
8\_2\_119 && 1440.0 & -5.0 & 2.0 & 0.0 & 0.94 & 1435.0\\
8\_5\_72 && 2764.0 & -31.2 & 3.24 & 0.0 & 0.92 & 2732.8\\
8\_2\_109 && 2313.0 & 0.0 & 3.5 & 0.0 & 0.94 & 2313.0\\
8\_4\_172 && 5235.0 & 0.0 & 2.65 & 0.0 & 0.89 & 5235.0\\
8\_5\_124 && 3559.0 & -5.0 & 2.4 & 0.0 & 0.86 & 3554.0\\
8\_5\_145 && 4682.0 & -3.7 & 1.66 & 0.0 & 0.9 & 4678.3\\
8\_5\_80 && 2745.0 & -64.9 & 2.66 & 1.0 & 0.89 & 2680.1\\
8\_6\_136 && 4316.0 & -4.5 & 3.1 & 0.0 & 0.92 & 4311.5\\
18\_10\_567 && 9380.0 & -11.8 & 2.64 & 4.3 & 0.9 & 9368.2\\
18\_9\_548 && 10192.0 & -3.4 & 2.79 & 4.8 & 0.91 & 10188.6\\
18\_4\_412 && 3430.0 & 0.0 & 1.25 & 11.1 & 0.93 & 3430.0\\
18\_6\_440 && 6050.0 & -13.0 & 2.67 & 11.5 & 0.89 & 6037.0\\
18\_4\_404 && 4957.0 & -25.0 & 2.83 & 1.1 & 0.91 & 4932.0\\
18\_8\_514 && 9453.0 & -99.0 & 1.89 & 4.2 & 0.87 & 9354.0\\
18\_10\_710 && 9100.0 & -55.3 & 2.24 & 1.9 & 0.88 & 9044.7\\
18\_8\_557 && 8643.0 & -1.4 & 2.48 & 2.6 & 0.91 & 8641.6\\
18\_4\_363 && 3400.0 & -73.0 & 1.75 & 6.5 & 0.91 & 3327.0\\
18\_9\_368 && 6841.0 & -41.0 & 2.18 & 10.4 & 0.9 & 6800.0\\
24\_45\_1072 && 31661.0 & -144.1 & 2.6 & 9.3 & 0.85 & 31516.9\\
24\_38\_906 && 32665.0 & -39.8 & 2.11 & 22.1 & 0.87 & 32625.2\\
24\_45\_621 && 32744.0 & -78.4 & 1.58 & 34.1 & 0.83 & 32665.6\\
24\_43\_767 && 35349.0 & -75.6 & 1.69 & 35.1 & 0.84 & 35273.4\\
24\_42\_749 && 36397.0 & -212.1 & 2.05 & 17.9 & 0.84 & 36184.9\\
24\_40\_825 && 38074.0 & -172.8 & 2.3 & 11.8 & 0.83 & 37901.2\\
24\_45\_1441 && 39311.0 & -111.1 & 1.9 & 8.0 & 0.84 & 39199.9\\
24\_45\_1076 && 36299.0 & -62.9 & 2.2 & 8.2 & 0.86 & 36236.1\\
24\_44\_610 && 32837.0 & -159.8 & 1.99 & 18.1 & 0.83 & 32677.2\\
24\_52\_662 && 36184.0 & -199.0 & 1.77 & 37.2 & 0.84 & 35985.0\\
72\_230\_2505 && 122491.0 & -409.5 & 2.33 & 57.6 & 0.82 & 122081.5\\
72\_211\_2438 && 129959.0 & -296.7 & 2.1 & 78.2 & 0.83 & 129662.3\\
72\_224\_2436 && 142132.0 & -333.3 & 1.93 & 103.1 & 0.81 & 141798.7\\
72\_211\_2400 && 144478.0 & -1488.3 & 2.0 & 95.1 & 0.82 & 142989.7\\
72\_216\_2027 && 128468.0 & -425.6 & 2.14 & 54.4 & 0.8 & 128042.4\\
72\_230\_3127 && 154910.0 & -854.4 & 2.38 & 40.6 & 0.82 & 154055.6\\
72\_213\_3244 && 145566.0 & -690.8 & 2.16 & 35.1 & 0.83 & 144875.2\\
72\_218\_2456 && 125318.0 & -248.3 & 2.03 & 68.3 & 0.82 & 125069.7\\
72\_234\_2543 && 146794.0 & -1123.7 & 2.21 & 75.9 & 0.82 & 145670.3\\
72\_180\_2258 && 92187.0 & -505.6 & 1.95 & 72.5 & 0.81 & 91681.4\\
\midrule
Average && 44751.4 & -202.16 & 2.32 & 23.55 & 0.87 & 44549.24\\
\bottomrule
\end{tabular}
\end{table}

\end{document}